\documentclass[floatfix,reprint,amsmath,amssymb,superscriptaddress,aps]{revtex4-2}
\usepackage{graphicx}  
\usepackage{dcolumn}   
\usepackage{bm}        
\usepackage{longtable}
\usepackage{amsmath}
\usepackage{tabularx}
\usepackage{tabulary}
\usepackage{makecell} 
\usepackage{gensymb}
\usepackage[pass,paperwidth=8.5in,paperheight=11in]{geometry}

\usepackage[linktocpage=true]{hyperref}
\usepackage{hyperref}
\usepackage{comment}
\usepackage{soul}
\usepackage{notes2bib}
\hypersetup{
  pdfnewwindow=true, 
  colorlinks=true,
  linkcolor=blue, 
  anchorcolor=blue,
  citecolor=blue, 
  filecolor=blue,
  menucolor=blue, 
  urlcolor=blue}

\usepackage[normalem]{ulem}
\usepackage{xcolor}
\usepackage{orcidlink}

\begin{document}
\newcommand\orangesout{\bgroup\markoverwith{\textcolor{orange}{\rule[0.5ex]{2pt}{0.4pt}}}\ULon}

\date{\today}

\title{Suppression of auxetic behavior in black phosphorus with sulfur substitution}

\author{Hayden Groeschel\,}
\affiliation{Department of Mechanical Engineering, University of Rochester, Rochester, New York 14627, USA}

\author{Arjyama Bordoloi\,\orcidlink{0009-0006-2760-3866}}
\affiliation{Department of Mechanical Engineering, University of Rochester, Rochester, New York 14627, USA}

\author{Sobhit Singh\,\orcidlink{0000-0002-5292-4235}}
\email{s.singh@rochester.edu}
\affiliation{Department of Mechanical Engineering, University of Rochester, Rochester, New York 14627, USA}
\affiliation{Materials Science Program, University of Rochester, Rochester, New York 14627, USA}

\begin{abstract}

Sulfur-doped black phosphorus (b-P) has recently emerged as a promising candidate for next-generation electronic and optoelectronic technologies owing to its enhanced environmental stability and tunable electronic properties. 
In this work, we systematically investigate the effects of sulfur substitution on the elastic, mechanical, and electronic properties of b-P, with a particular focus on its auxetic behavior (i.e., negative Poisson’s ratio), using first-principles density-functional theory calculations. 
Our results unveil the fundamental origin of the intrinsic auxetic response in pristine b-P and elucidate how sulfur incorporation alters this behavior. We find that sulfur atoms distort the characteristic bow-tie structural motif responsible for the negative Poisson’s ratio in b-P, thereby suppressing the in-plane auxeticity.
Moreover, the resulting charge redistribution also effectively quenches the out-of-plane auxetic response of b-P. 
With increasing sulfur content, the bulk modulus and Poisson’s ratio increase, whereas the Young’s modulus, shear modulus, and Debye temperature decrease. Additionally, sulfur substitution suppresses the semiconducting properties of b-P, giving rise to metallicity. 
These findings highlight that, although sulfur substitution enhances the environmental stability of b-P, it also substantially modifies its elastic and mechanical properties, particularly the auxetic behavior, which is an important consideration in the design of nanoscale electronic devices.

\end{abstract}

\maketitle

\section{Introduction}

With the recent surge of interest in layered 2D materials, black phosphorus (b-P) and its monolayer counterpart, black phosphorene, have become the focus of extensive research~\cite{Du_Nano_Letters_2016,Du_journal_applied_phys_2010,Mu_Mat_today_phy_2019,Appalakondaiah_PhysRevB_2012}. These materials are particularly attractive due to their auxetic behavior, characterized by an intrinsic negative Poisson’s ratio~\cite{Du_Nano_Letters_2016}, as well as their tunable electronic properties and thickness-dependent
charge-carrier mobility~\cite{Li_Nature_nanotech_2014, Tran_PhysRevB_2014, Huang_Applied_Physics_Letters_2016}. Black phosphorus exhibits a direct bandgap ranging from $\sim$\,0.33-0.35\,eV in bulk to about 2\,eV in the monolayer, with the electronic bandgap strongly dependent on the number of layers~\cite{Warschauer_Journal_AP_1963, Tran_PhysRevB_2014, Li_Nature_nanotech_2014, Huang_Applied_Physics_Letters_2016, Guan_PhysRevB_2016}. This tunable bandgap enables few-layer b-P to be readily switched between ON and OFF conduction states, making it a highly promising layered material for optoelectronic devices and field-effect transistors~\cite{Ponaga_PhysRevB_2022,Yi_small_methods_2019,Debnath_small_methods_2018}. 
Moreover, surface functionalization can substantially tailor its intrinsic properties, further broadening its potential for real-world applications~\cite{Xiang_nature_comms_2015,Koenig_nano_letters_2016,Xu_small_2016,CHEN_carbon_2017,Wu_Angewandte_chemie_international_2021}. Notably, superconductivity has been reported in b-P under high pressure~\cite{Li_national_academy_sciences_2018}.

\begin{figure*}
\centering
\includegraphics[width=0.75\textwidth]{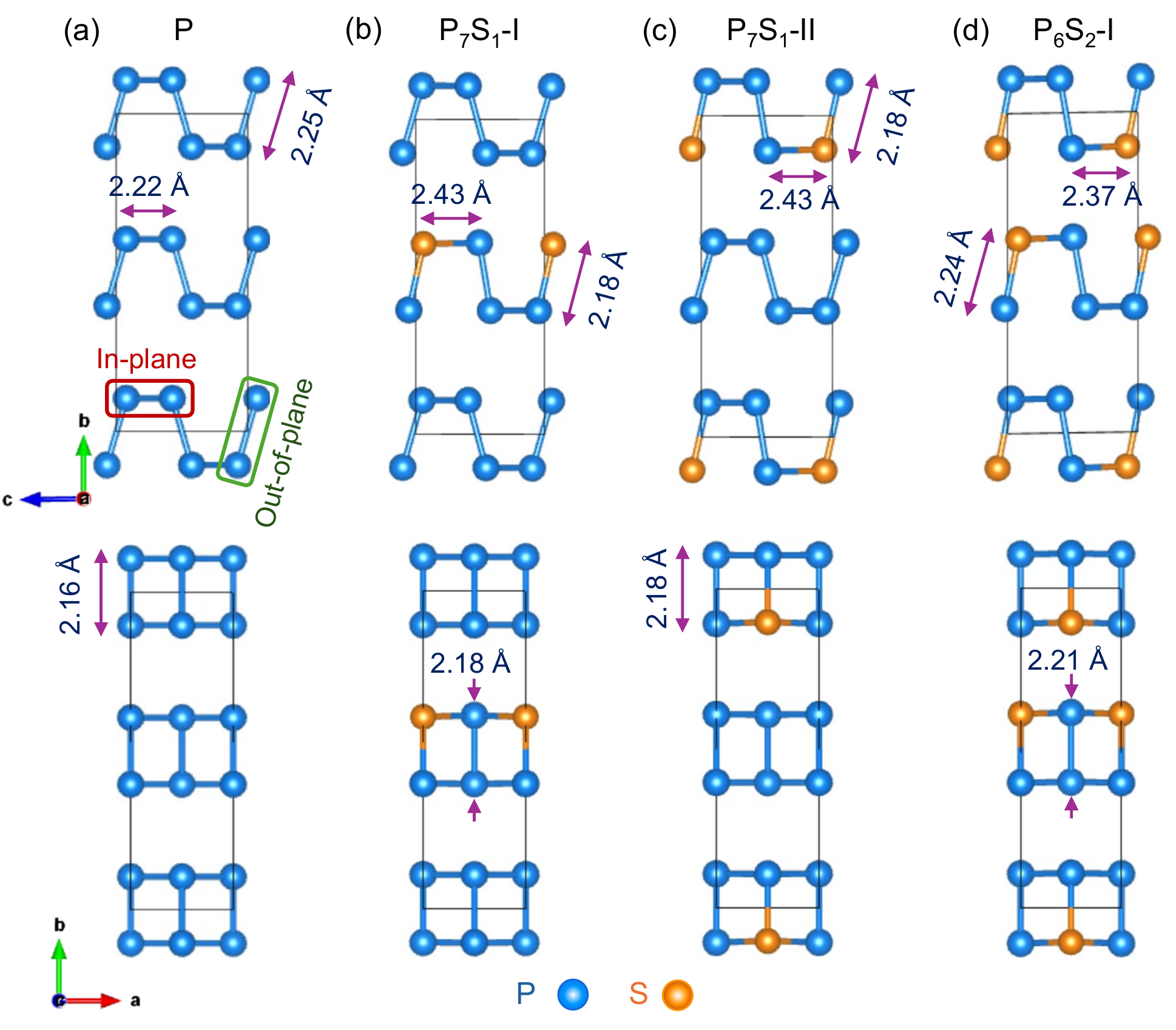}
\caption{Crystal structures of bulk black P$_{1-x}$S$_{x}$. Phosphorus atoms are shown in blue, and sulfur atoms are shown in orange.}
\label{fig: POSCAR}
\end{figure*}

Among the diverse range of promising applications of b-P, one particularly exciting prospect lies in its suitability for next-generation field-effect transistors (FETs)~\cite{Li_Nature_nanotech_2014}, which is a research direction of increasing importance as silicon-based technologies approach their fundamental physical scaling limits~\cite{he_high_k_gate_dielectrics_2012}.
b-P can be realized as an ultra-thin material with electronic properties and charge-transport characteristics comparable to those of silicon-based FETs~\cite{Li_Nature_nanotech_2014}, which makes it an attractive candidate for future nanoscale electronic devices. 
However, instability under ambient conditions has been observed in pristine b-P~\cite{vanDruenen_AdvMatInterfaces_2020,Island_2D_materials_2015}, leading to a search for effective stabilization strategies~\cite{Lv_ACS_applied_Materials_2018,Sarswat_electrochem_society_2016,Yang_advanced_materials_2016,Doganov_nature_comms_2015,Wan_nanotechnology_2015}. 

Lv \textit{et al}.~\cite{Lv_ACS_applied_Materials_2018} proposed sulfur (S) doping as an effective strategy to suppress b-P degradation. Sulfur doping enhances the ambient stability of few-layer b-P FETs by inhibiting oxidation and is more effective than other chalcogen elements, such as Te~\cite{Lv_ACS_applied_Materials_2018, song_Adv_functional_materials_2019}. Although the substitution of S has been investigated to improve the environmental stability of b-P, its effects on other properties, particularly electronic, elastic, and mechanical behavior, including the unique auxetic behavior of b-P, are not yet explored. A comprehensive understanding of these properties is essential, as they are critical for the practical design and reliability of b-P-based devices.





To address this knowledge gap, we systematically investigate the effect of sulfur substitution on the electronic, elastic, and mechanical properties of black phosphorus using first-principles density functional theory (DFT) calculations. 
Our results reveal that sulfur, being more electronegative than phosphorus, distorts the characteristic bow-tie structural motif of pristine b-P, which is responsible for its intrinsic negative Poisson’s ratio, thus suppressing the in-plane auxeticity. 
Furthermore, the accompanying charge redistribution quenches the out-of-plane auxetic response and significantly alters other elastic and mechanical properties.
Although S-substituted b-P remains mechanically stable up to a sulfur content of 25\,\%, a further increase in sulfur concentration induces mechanical instability. Additionally, sulfur substitution transforms semiconducting b-P into a metallic system. 
Our findings indicate that, although S doping enhances the environmental stability of b-P, there exists a practical limit to sulfur incorporation, beyond which mechanical instability may render the material unsuitable for device applications. 

\begin{table}[!b]
\centering
\renewcommand*{\arraystretch}{1.2}
\caption{Optimized lattice parameters and space groups of black P$_{1-x}$S$_{x}$ (bulk) calculated using PBEsol functional. Percentage errors of lattice parameters with respect to the experimental data from Ref.~\cite{Cartz_Journal_Chemical_Physics_1979} are shown in parentheses. } 
\label{tab:lattice parameters}
\begin{tabularx}{\columnwidth}{l@{\hspace{0.5em}}|*{4}{>{\centering\arraybackslash}X}}
\hline
Composition & a (\r{A}) & b (\r{A}) & c (\r{A}) & Space Group \\
\hline
b-P (PBEsol) & 3.314 (0.03\%) & 10.322 (1.44\%) & 4.217 (3.59\%) & Cmce\,(64) \\
\makecell[l]{b-P (PBEsol+D3)} & 3.324 (0.33\%) & 10.162 (2.97\%) & 4.101 (6.24\%) & Cmce\,(64) \\
Expt.~\cite{Cartz_Journal_Chemical_Physics_1979} & 3.313 & 10.473 & 4.374 & Cmce\,(64) \\
\hline
P$_7$S$_1$-I & 3.263 & 10.157 & 4.174 & Pm\,(6) \\
P$_7$S$_1$-II & 3.264 & 10.157 & 4.173 & Pm\,(6) \\
P$_6$S$_2$-I & 3.203 & 10.164 & 4.112 & P2$_1$/m\,(11) \\
\hline
\end{tabularx}
\end{table}

\begin{table*}[tbh]
\renewcommand*{\arraystretch}{1.25}
\caption{Elastic constants (C$_{ij}$)  for different P$_{1-x}$S$_{x}$ configurations with varying sulfur concentrations, computed using PBEsol in units of GPa. Experimental data for pristine b-P, from Ref.~\cite{Ponaga_PhysRevB_2022}, are reported for comparision. 
}

\label{tab:elastic constants}
\begin{tabularx}{\textwidth}{l|*{9}{>{\centering\arraybackslash}X}}
\hline
Composition~~~& ~~C$_{11}$ ~~&~~ C$_{22}$~~ & ~~C$_{33}$~~ & ~~C$_{44}$~~ & ~~C$_{55}$ ~~ & ~~C$_{66}$ ~~ & ~~ C$_{12}$ ~~ & ~~ C$_{13}$~~ & ~~ C$_{23}$~~\\
\hline\
P & 188.5 & 55.0 & 45.6 & 6.5 & 77.2 & 17.4 & 4.2 & 49.0 & -5.6\\
Expt.~\cite{Ponaga_PhysRevB_2022} & 193 & 53 & 61 & 7 & 72 & 27 & 3 & 50 & -7\\
\hline\
P$_7$S$_1$-I & 173.0 & 54.8 & 40.2 & 7.0 & 61.6 & 15.5 & 8.3 & 49.3 & 7.1\\
P$_7$S$_1$-II & 173.0 & 54.9 & 40.3 & 6.9 & 61.7 & 15.6 & 8.2 & 49.3 & 6.9\\
P$_6$S$_2$-I & 176.6 & 57.1 & 62.1 & 14.3 & 61.7 & 4.2 & 6.3 & 66.1 & 13.6\\

\hline
\end{tabularx}
\end{table*}

\section{Computational Details}
First-principles density functional theory (DFT) calculations were performed using the projector augmented wave (PAW) method, as implemented in the Vienna Ab initio Simulation Package (VASP)~\cite{Kresse96a,Kresse96b,KressePAW, Blochl94}. 
Exchange-correlation effects were treated using the semilocal Perdew-Burke-Ernzerhof functional for solids (PBEsol)~\cite{PBEsol}. 
Inclusion of van der Waals (vdW) interactions via the DFT-D3 scheme~\cite{Grimme_DFT_D3_2010,Grimme_DFT_D3_2011} resulted in pronounced overbinding of the optimized lattice parameters of b-P, as evident from Table~\ref{tab:lattice parameters}. 
This behavior arises because the semilocal PBEsol functional already captures the weak interlayer vdW interactions in b-P reasonably well. Therefore, all subsequent calculations were performed without additional vdW corrections.
%

A plane-wave kinetic energy cutoff of 450\,eV was employed, and the electronic self-consistent field  convergence criterion was set to 10$^{-7}$\,eV. Full relaxation of ionic positions, cell shape, and cell volume was carried out until the residual Hellmann-Feynman forces on each atom were less than 10$^{-3}$\,eV/\r{A}. A Monkhorst–Pack k-point mesh of 12 $\times$ 6 $\times$ 12 was used for Brillouin zone sampling. In the PAW pseudopotentials, five valence electrons were considered for phosphorus (3s$^{2}$3p$^{3}$) and six for sulfur (3s$^{2}$3p$^{4}$). The elastic tensor $C_{ij}$ was computed via the stress-strain method as implemented in VASP, and subsequent analysis of elastic properties including the elastic moduli, elastic wave velocities and Debye temperature was performed using the {\sc MechElastic} Python package~\cite{mechelastic, SinghPRB2018_meche}.
The spatial dependence of the Poisson’s ratio was analyzed using the ELATE software~\cite{Gaillac_2016}, while the electron localization function (ELF) distribution was visualized using VESTA~\cite{VESTA}. The screened hybrid functional HSE06~\cite{HSE06_1,HSE06_2} was used to compute the electronic density of states for the studied systems. 

\section{Results and Discussions}
 \subsection{Crystal Structures}

Figure~\ref{fig: POSCAR} illustrates the crystal structure of pristine and sulfur-substituted black phosphorus (b-P) from two different viewing directions. Pristine b-P adopts a puckered layered structure with space group $Cmce$ (no.~64), as shown in Fig.~\ref{fig: POSCAR}(a). 
The DFT-predicted optimized lattice parameters computed using PBEsol functional~\cite{PBEsol} are in excellent agreement with experimental values as listed in Table~\ref{tab:lattice parameters}. 
Notably, inclusion of vdW corrections using the DFT-D3 scheme led to a pronounced overbinding of the optimized lattice parameters, resulting in larger deviations from the experimental values. Therefore, vdW corrections were excluded from all subsequent calculations.

To study the effect of sulfur substitution, four symmetry-inequivalent P sites were selected, yielding six distinct configurations: two with a single S atom substitution (P$_7$S$_1$-I and P$_7$S$_1$-II) in one of two phosphorene layers of the bulk structure [see Figs.~\ref{fig: POSCAR}(b) and (c)], 
one with one S atom in each phosphorene of the bulk structure (P$_6$S$_2$-I), 
two with two S atoms in the same phosphorene layer (P$_6$S$_2$-II and P$_6$S$_2$-III), and one with two S atoms per layer, corresponding to a 50$\%$ S substitution (P$_4$S$_4$).
The structures of P$_6$S$_2$-II, P$_6$S$_2$-III, and P$_4$S$_4$ are provided in the Supplementary Figure S1~\cite{SM}.

\begin{table*}[tbh]
\renewcommand*{\arraystretch}{1.25}
\caption{Elastic properties of P$_{1-x}$S$_{x}$, including the bulk modulus K (GPa), shear modulus G (GPa), Young’s modulus E (GPa), average Poisson’s ratio $\nu$, longitudinal v$_l$, transverse v$_t$, and average v$_m$ elastic wave velocities (km/s), along with the Debye temperature $\Theta_{Debye}$ (K). Experimental values for K and G are reported in Ref.~\cite{Ponaga_PhysRevB_2022}, while E and $\nu$ are calculated using standard relationships with K and G.}
\label{tab: mechanical properties}
\begin{tabularx}{\textwidth}{l|*{8}{>{\centering\arraybackslash}X}}
\hline
Composition~~~& ~~ K (GPa) ~~&~~ G (GPa) ~~ & ~~E (GPa)~~ & ~~ $\nu$ ~~ & ~~$v_{l}$\,(km/s)~~ & ~~$v_{t}$\,(km/s)~~ & ~~ $v_{m}$\,(km/s)~~ & ~~ $\Theta_{\mathrm{Debye}}$\,(K)~~\\
\hline\
P & 32.3 & 26.2 & 61.8 & 0.187 & 4.855 & 3.030 & 3.338 & 379.0\\
Expt.~\cite{Ponaga_PhysRevB_2022} & 34 & 26 & 62 & 0.195 & - & - & - & -\\
P$_7$S$_1$-I & 35.2 & 22.7 & 55.9 & 0.240 & 4.678 & 2.754 & 3.052 & 351.4\\
P$_7$S$_1$-II & 35.1 & 22.7 & 56.0 & 0.240 & 4.679 & 2.756 & 3.054 & 351.6\\
P$_6$S$_2$-I & 44.2 & 20.9 & 53.8 & 0.306 & 4.821 & 2.598 & 2.900 & 337.6\\

\hline
\end{tabularx}
\end{table*}

\begin{figure}[!!b]
\centering
\includegraphics[width=9cm]{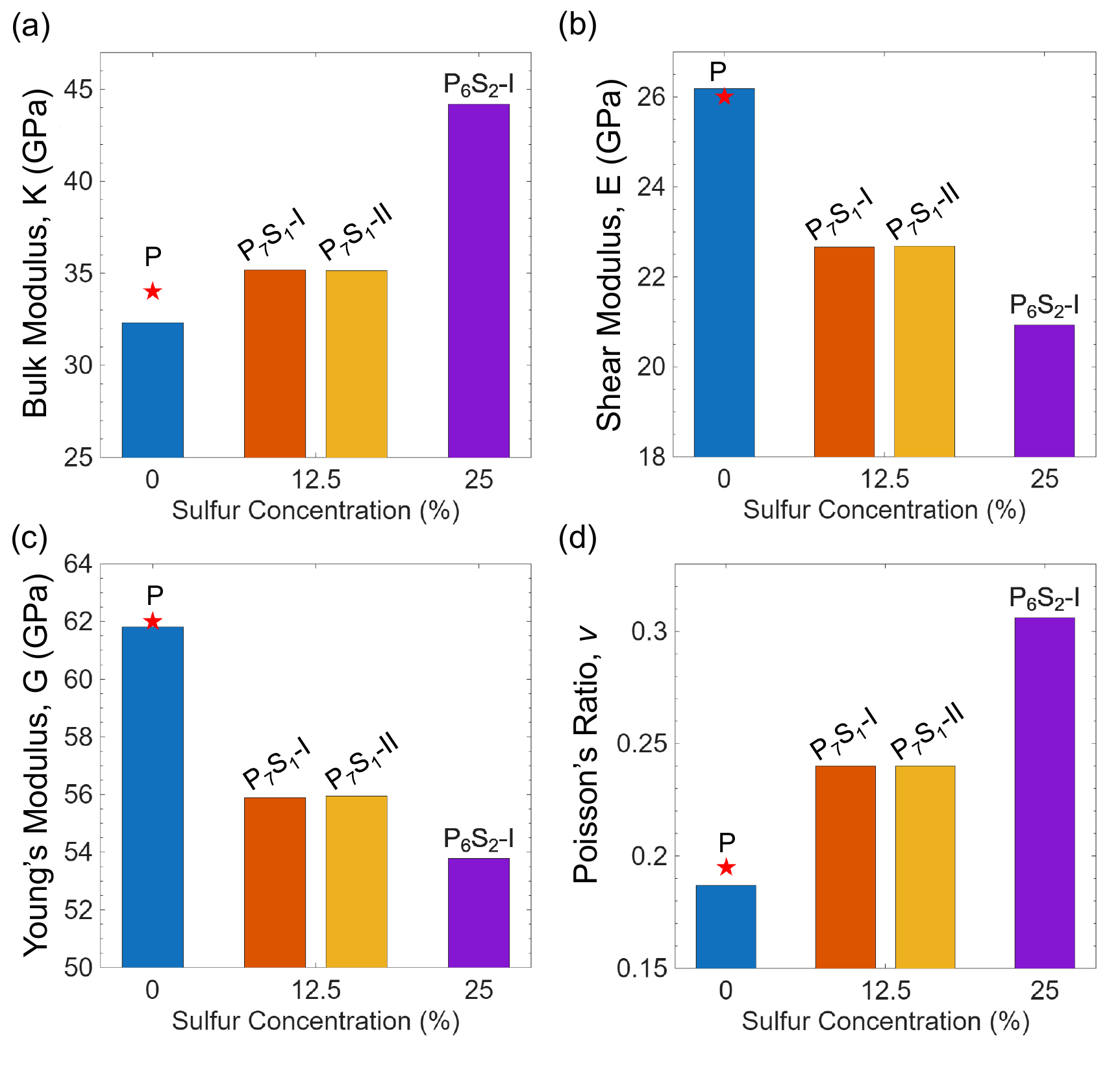}
\caption{Elastic properties of P$_{1-x}$S$_{x}$: (a) bulk modulus K\,(GPa), (b) shear modulus E\,(GPa), (c) Young's modulus G\,(GPa), and (d) average Poisson’s ratio $\nu$. Experimental values for K and G are reported from Ref.~\cite{Ponaga_PhysRevB_2022}, while E and $\nu$ are calculated using standard relationships with K and G. Experimental values are shown as red stars.
}

\label{fig: elastic}
\end{figure}

\begin{figure*}[tbh]
\centering
\includegraphics[width=1\textwidth]{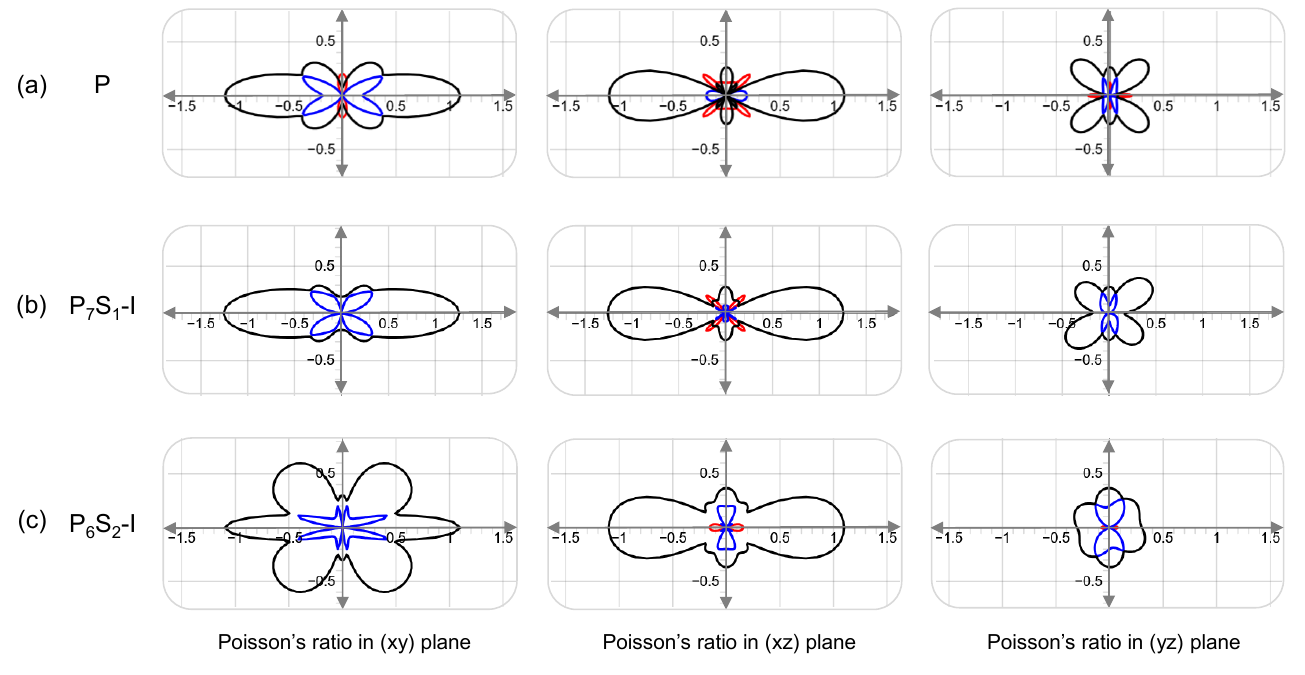}
\caption{Spatial dependence of Poisson's ratio for pure black phosphorus, P$_{7}$S$_{1}$-I, and P$_{6}$S$_{2}$-I, respectively. All plots were generated using the ELATE software~\cite{Gaillac_2016}. Blue lines represent the positive Poisson's ratio, and red lines represent the negative Poisson's ratio.}
\label{fig: poisson}
\end{figure*}

For the single-S substitution, the resulting P$_7$S$_1$-I and P$_7$S$_1$-II configurations crystallize in the $Pm$ symmetry (space group no.~6) and exhibit nearly similar mechanical and electronic properties, as discussed below. 
P$_6$S$_2$\,-I, on the other hand, optimizes into $P2_1/m$ symmetry (space group no.~11) [Fig.~\ref{fig: POSCAR}(d)]. Notably, the other two configurations with 25$\%$ S substitution (P$_6$S$_2$-II and P$_6$S$_2$-III), as well as the configuration with 50$\%$ S substitution\,(P$_4$S$_4$) undergo severe structural distortions that disrupt the layered framework; therefore, their properties are reported in the Supplementary Material~\cite{SM} rather than in the main text. The DFT (PBEsol) optimized lattice parameters along with the space groups of the studied configurations are summarized in Table~\ref{tab:lattice parameters}.

Sulfur substitution has minimal impact on the lattice parameters, with variations not exceeding 3.6\%. As shown in Figure~\ref{fig: POSCAR}, each substituted sulfur atom forms three bonds with neighboring P atoms: two in-plane P–S bonds and one out-of-plane P–S bond per phosphorene. For both P$_7$S$_1$ configurations, we see an elongation in the in-plane P–S bonds from 2.22\,Å in pristine b-P to 2.43\,Å, while the out-of-plane bond shortens from 2.25\,Å to 2.18\,Å. In contrast, for P$_6$S$_2$-I, the in-plane P–S bonds increase to 2.37\,Å, with a slight decrease in the out-of-plane bond to 2.24\,Å. 
Furthermore, the layer thickness increases from 2.16\,Å in pristine b-P to 2.18\,Å for P$_7$S$_1$ and 2.21\,Å for P$_6$S$_2$-I.


\begin{figure*}
\centering
\includegraphics[width=1\textwidth]{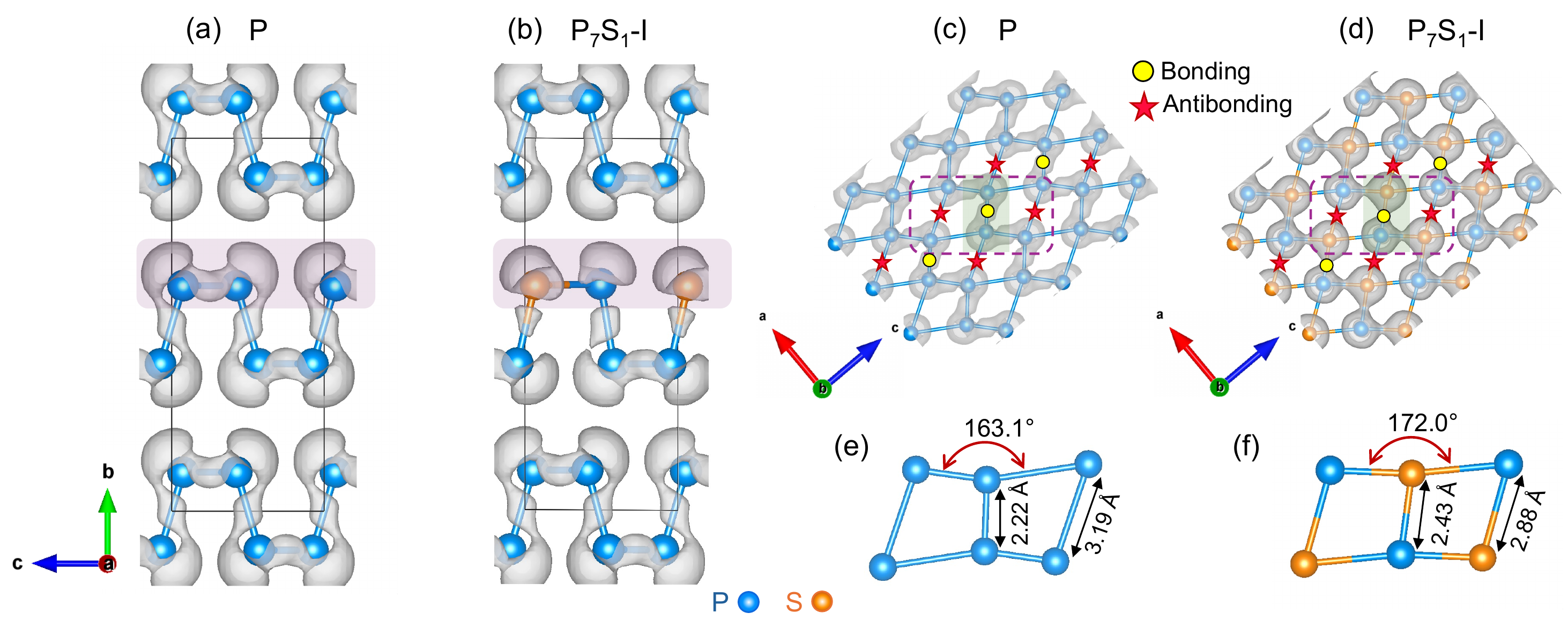}
\caption{Electron localization function (ELF) distribution (gray) for (a) b-P and (b) P$_7$S$_1$-I.
(c)–(d) Top view of the ELF for the upper P–P (P–S) plane [highlighted in pink in (a) and (b)] in a single monolayer of pristine b-P and P$_7$S$_1$-I, respectively, illustrating the characteristic bow-shaped structural configuration responsible for the in-plane auxeticity in the xz-plane of b-P. (e)–(f) Corresponding P–P–P and P–S–P bond angles, along with P–P and P–S bond lengths, showing how S substitution distorts the bow-type configuration. Yellow circles represent bonding interactions, while red stars denote antibonding behavior.}

\label{fig: auxeticity}
\end{figure*}

\subsection{Elastic and Mechanical Properties}

The elastic moduli of a material, namely the bulk modulus (K), shear modulus (G), Young's modulus (E), and Poisson's ratio ($\nu$), are key parameters for understanding its mechanical behavior. These quantities describe how a material responds under different deformation conditions and are therefore essential when considering a material for practical applications. A general approach to determine these values is by use of a stiffness tensor, which is obtained using the generalized stress-strain Hooke's law~\cite{nye1985physical}. This can be expressed as,
\begin{equation}
\sigma_{ij} = C_{ijkl} \epsilon_{kl},
\end{equation}
where ${\sigma}_{ij}$ and $\epsilon_{kl}$ represent homogeneous second-rank stress and strain tensors, respectively, and \(C_{ijkl}\) represents the fourth-rank elastic stiffness tensor. By considering the symmetries of the stress and strain tensors, the number of independent \(C_{ijkl}\) components can be reduced from 81 to 36. Further, considering the crystal symmetry of the studied systems, only 9 independent coefficients are required for a complete analysis of mechanical properties.

Table~\ref{tab:elastic constants} presents the elastic constants for all studied configurations, along with a comparison to the experimental values for pristine b-P. All structures satisfy the Born-Huang mechanical stability criteria corresponding to their respective space group symmetries, i.e., orthorhombic or monoclinic~\cite{mechelastic}, which require the stiffness matrix $C_{ij}$ to be symmetric and all eigenvalues to be positive. Pristine b-P satisfies the Born-Huang criteria for an orthorhombic structure, which are~\cite{mechelastic}: C$_{11}$\,$>$\,0, C$_{11}$C$_{22}$\,$>$\,C$_{12}$$^{2}$, C$_{11}$C$_{22}$C$_{33}$\,+\,2$C_{12}$C$_{13}$C$_{23}$\,-\,C$_{11}$C$_{23}$$^{2}$\,-\,C$_{22}$C$_{13}$$^{2}$\,-\,C$_{33}$C$_{12}$$^{2}$\,$>$ 0, C$_{44}$\,$>$\,0, C$_{55}$\,$>$\,0, C$_{66}$\,$>$\,0. 
On the other hand, the P$_7$S$_1$ and P$_6$S$_2$ configurations satisfy the Born-Huang stability criteria for monoclinic structures. In contrast, the 50\% S-substituted case violates these criteria, indicating mechanical instability, as discussed in the Supplementary Materials~\cite{SM}. Notably, both P$_7$S$_1$ configurations exhibit almost identical values of elastic constants.

Once the $C_{ij}$ constants are determined, the elastic moduli can be derived from their relationships with the elastic constants. The bulk modulus and shear modulus are calculated using the Voigt-Reuss-Hill averaging scheme~\cite{hill1952elastic}, while the Young's modulus and Poisson's ratio $\nu$ were derived from their standard relationships with K and G~\cite{mechelastic}.

Table~\ref{tab: mechanical properties} summarizes the calculated elastic moduli, while Figure~\ref{fig: elastic} presents them as a bar graph, including the experimental values estimated for pristine b-P for comparison. The reported experimental values for K and G are taken from Ref.~\cite{Ponaga_PhysRevB_2022}, and values for E and $\nu$ are calculated using standard relationships with K and G~\cite{mechelastic}. 
%
As shown in Figure~\ref{fig: elastic}, both the bulk modulus and Poisson’s ratio increase with increasing sulfur concentration. This trend is attributed to the reduction in all lattice parameters upon sulfur substitution (see Table~\ref{tab:lattice parameters}). 
In contrast, the Young’s and shear moduli decrease with increasing sulfur concentration, primarily due to the elongation of in-plane P–S bonds.


\textbf{Suppression of auxetic behavior:}
One of the primary objectives of this study is to investigate how sulfur substitution influences the distinctive auxetic behavior of b-P. 
Figure~\ref{fig: poisson} shows the spatial distribution of the Poisson’s ratio for pristine b-P and two sulfur-substituted configurations (P$_7$S$_1$-I and P$_6$S$_2$-II), computed using the ELATE software~\cite{Gaillac_2016}. Pristine b-P exhibits negative Poisson’s ratios both along the out-of-plane ($\hat{y}$ // ${\bf b}$) direction and along the diagonal directions of the xz-plane (i.e., ${\bf a}$-${\bf c}$ plane), as indicated by the red butterfly-shaped pattern in Figure~\ref{fig: poisson}(a). Upon sulfur substitution, out-of-plane auxeticity is completely suppressed, and the in-plane auxetic behavior is significantly reduced, as shown in Figures~\ref{fig: poisson}(b) and~\ref{fig: poisson}(c).

Consistent with earlier studies~\cite{Jiang_Nature_comm_2014}, the out-of-plane auxeticity in b-P originates from its puckered structure, which behaves like a pair of orthogonally coupled hinges. 
To gain deeper insight into the origin of in-plane auxeticity, we analyze the atomic structure and electron localization function (ELF) of b-P in detail, as shown in Fig.~\ref{fig: auxeticity}(c). The ELF is projected onto the lattice of P atoms within one of the two layers of the puckered phosphorene structure in bulk b-P, as indicated by the pink-shaded rectangle in Figs.~\ref{fig: auxeticity}(a,b). We note that both the top and bottom P-P layers are structurally identical in phosphorene, as illustrated in Fig.~\ref{fig: auxeticity}(c).

Interestingly, we observe a periodic pattern of bonding and antibonding orbital-like distributions along the alternating P-P bonds within the xz-plane, as marked in Fig.~\ref{fig: auxeticity}(c). 
These P-P bonds form the characteristic bow-tie-like structural motif of b-P, featuring a hinge-like configuration of P-P bonds as illustrated in Fig.~\ref{fig: auxeticity}(e). 
We note such a bow-tie-like structural motif with periodically alternating bonding and antibonding P-P covalent bonds is the fundamental origin of in-plane auxetic behavior in b-P.
Under tensile strain, the bow-tie-shaped units flatten toward a more rectangular configuration, leading to an increase in the transverse dimension rather than contraction, thereby resulting in a negative Poisson’s ratio.


When one P atom is substituted with one S atom, the higher electronegativity of sulfur draws electron density away from the bonding region, as reflected in the ELF profiles shown in Figs.~\ref{fig: auxeticity}(c) and \ref{fig: auxeticity}(d), 
thus introducing some ionic character in the 
P-S bond and elongating it by $\sim$\,10\% to 2.43~\AA, as compared to 2.22~\AA~for the P–P bond in pristine b-P. 
This redistribution of charge increases the P-S-P bond angle to 172$\degree$ from the corresponding P-P-P angle of 163.1$\degree$ in b-P [Figs.~\ref{fig: auxeticity}(e) and (f)], thereby effectively flattening the local bow-tie structure and reducing its hinge flexibility. This attenuated hinge-like motion suppresses the in-plane auxeticity.

Furthermore, the electron withdrawal by sulfur enhances charge localization near the layer boundaries, thereby strengthening interlayer coupling. This increased interlayer stiffness makes the layers more resistant to deformation, contributing to both the higher bulk modulus and the complete suppression of out-of-plane auxeticity. As shown in Figure~\ref{fig: poisson}(a), pristine b-P demonstrates auxetic behavior under tensile loading along the out-of-plane y-direction, whereas this feature is nearly absent in the sulfur-substituted structures, as depicted in Figures~\ref{fig: poisson}(b) and~\ref{fig: poisson}(c).

\begin{figure}[b]
\centering
\includegraphics[width=9cm]{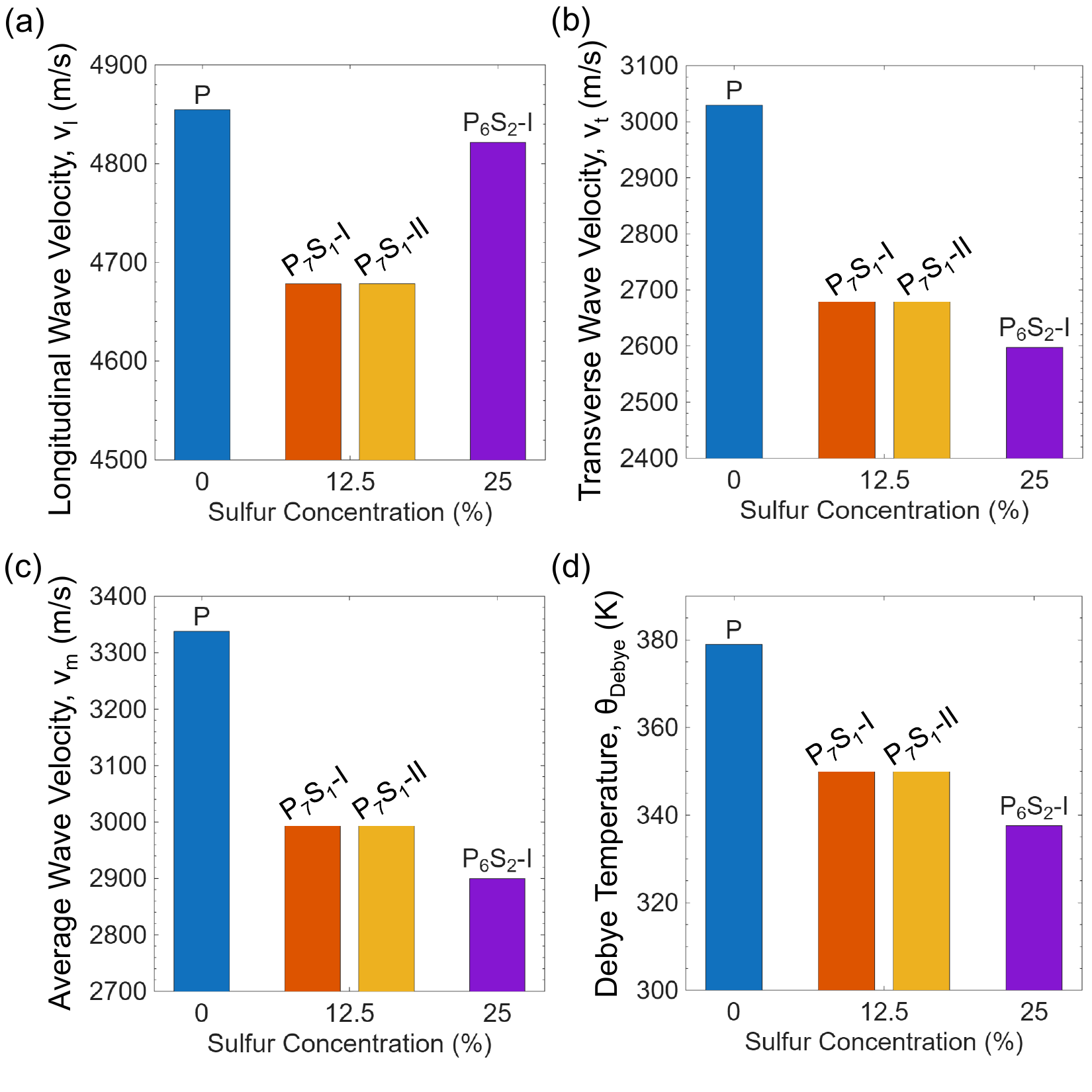}
\caption{Elastic wave velocities and Debye temperature of P$_{1-x}$S$_{x}$ configurations with varying sulfur concentration.} 
\label{fig: waves}
\end{figure}

\subsection{Elastic wave velocities}

Next, we analyze the propagation of elastic waves in the studied material. 
The longitudinal ($v$$_{l}$) and transverse ($v$$_{t}$) wave velocities are calculated using the following relations: 
\begin{equation}
v_{l}=\sqrt{\frac{3K+4G}{3\rho}},
\end{equation}
\begin{equation}
v_{t}=\sqrt{\frac{G}{\rho}},
\end{equation}
and the average ($v$$_{m}$) wave velocity is calculated using the Voigt-averaging formula as follows:
\begin{equation}
v_{m}=\bigg[\frac{1}{3}\bigg(\frac{2}{v_t^3}+\frac{1}{v_l^3}\bigg)\bigg]^{-1/3},
\end{equation}
where $\rho$ refers to the density of the material. All values were computed using the {\sc MechElastic} software~\cite{mechelastic} and are reported in Table~\ref{tab: mechanical properties}. Figures~\ref{fig: waves}(a)–(c) represent the elastic wave velocities for the different sulfur-substituted configurations. As expected, the transverse wave velocity, which is proportional to $\sqrt{G}$, decreases monotonically. Interestingly, the longitudinal wave velocity of P$_6$S$_2$-I is nearly identical to that of pristine b-P. This can be attributed to the significant increase in bulk modulus for the 25\% sulfur-substituted case. 

\begin{figure}[tbh]
\centering
\includegraphics[width=0.95\columnwidth]{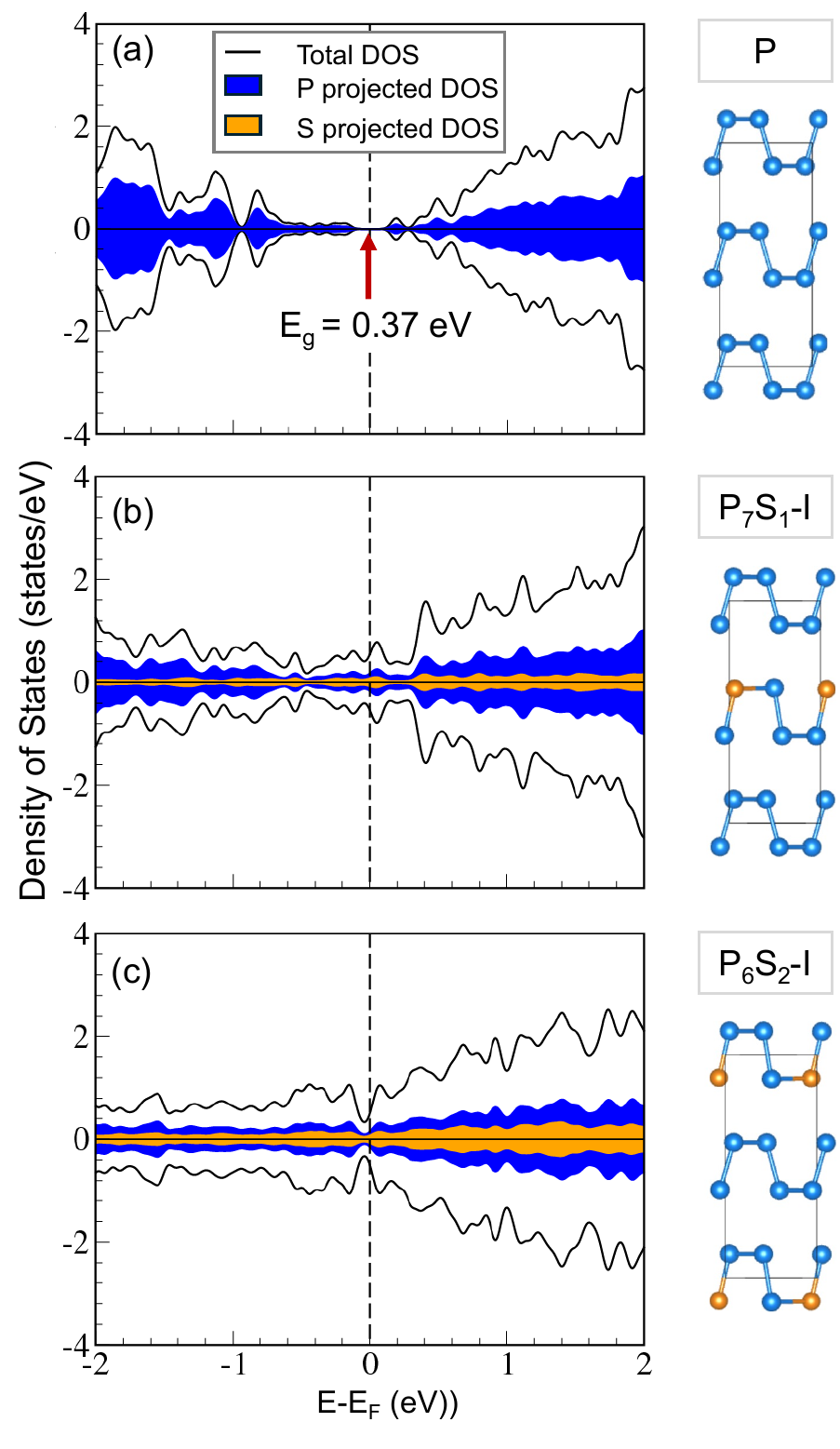}
\caption{Density of states for three cases of P$_{1-x}$S$_{x}$, where E$_{g}$ represents the band gap. The vertical dashed line represents the Fermi level (E$_{F}$).}
\label{fig: DOS}
\end{figure}

Finally, we calculate the Debye temperature ($\Theta_D$), which is an important thermodynamic parameter that marks the boundary between quantum and classical vibration regimes (i.e., below $\Theta_D$, lattice vibrations are quantized, while above it, they can be approximated classically). 
Debye temperature can be estimated using the average elastic wave velocity $v$$_{m}$ and the density of the material $\rho$~\cite{mechelastic}. 
It is calculated using the following formula:
\begin{equation}
\Theta_{D}=\frac{h}{k_{B}}\bigg{[}\frac{3q}{4\pi}\frac{N\rho}{M}\bigg{]}^{1/3}v_m,
\end{equation}
where $h$ is the Planck's constant, $k_{B}$ is the Boltzmann's constant, $q$ is the total number of atoms in the cell, $N$ is the Avogadro's number, and $M$ is the molecular weight. 
It should be noted that the $\Theta_D$ estimated from the elastic constants closely matches the $\Theta_D$ obtained from the low-temperature specific heat measurements, since at low temperatures only acoustic phonons contribute significantly to the specific heat~\cite{anderson_Jounal_of-phy_chem_of_solids_1963}.
The calculated values are reported in Table~\ref{tab: mechanical properties}, and Figure~\ref{fig: waves}(d) presents them as a bar graph. As expected, the Debye temperature decreases monotonically, consistent with the decreasing trend of the average elastic wave velocity.

\subsection{Electronic Properties}

For a complete understanding of the effect of sulfur substitution in b-P, we compute the electronic density of states (DOS) for the studied configurations to examine its influence on the electronic properties. Figure~\ref{fig: DOS} shows the DOS for these configurations, projected onto both P and S atoms, with the dashed line representing the Fermi level. 
Pristine b-P exhibits semiconducting behavior with a computed HSE06 band gap of 0.37\,eV, which is in good agreement with previous experimental reports~\cite{Warschauer_Journal_AP_1963, Tran_PhysRevB_2014, Guan_PhysRevB_2016}. Upon sulfur substitution, the system shows metallic behavior, as indicated by the nonzero DOS at the Fermi level. Both P and S atoms have dominant contributions to the DOS near the Fermi level, 
indicating strong hybridization between their $3p$ valence electronic orbitals.

\section{Conclusion}
This work investigates the effect of sulfur substitution on the elastic, mechanical, and electronic properties of black phosphorus. Sulfur incorporation significantly modifies the mechanical and elastic behavior of b-P, notably suppressing its auxetic behavior. 
Sulfur, being more electronegative than phosphorus, distorts the characteristic bow-tie structural motif of pristine b-P, which is responsible for its negative Poisson’s ratio, thus suppressing in-plane auxeticity. 
The out-of-plane auxetic behavior also vanishes due to the associated charge redistribution. 
With increasing sulfur content, the bulk modulus and Poisson's ratio increase, whereas the Young's modulus and shear modulus decrease. The Debye temperature also decreases with increasing sulfur substitution. Additionally, higher sulfur concentrations transform b-P from a semiconductor to a metallic system. Although sulfur doping has been previously explored as a strategy to stabilize b-P, our study highlights its significant impact on elastic, mechanical and electronic properties, particularly on the auxetic behavior, which is 
a crucial factor for the rational design of practical flexible and nanoscale electronic devices.



\section*{Acknowledgements}
Authors acknowledge support from the U.S.~Department of Energy, Office of Science, Office of Fusion Energy Sciences, Quantum Information Science program under Award No.~DE-SC-0020340.
Authors also acknowledge support from the Furth Research Fund at the University of Rochester. 
Authors thank the Pittsburgh Supercomputer Center (Bridges2) supported by the Advanced Cyberinfrastructure Coordination Ecosystem: Services \& Support (ACCESS) program, which is supported by National Science Foundation grants \#2138259, \#2138286, \#2138307, \#2137603, and \#2138296.  


\bibliography{main.bbl}

\begin{thebibliography}{46}%
\makeatletter
\providecommand \@ifxundefined [1]{%
 \@ifx{#1\undefined}
}%
\providecommand \@ifnum [1]{%
 \ifnum #1\expandafter \@firstoftwo
 \else \expandafter \@secondoftwo
 \fi
}%
\providecommand \@ifx [1]{%
 \ifx #1\expandafter \@firstoftwo
 \else \expandafter \@secondoftwo
 \fi
}%
\providecommand \natexlab [1]{#1}%
\providecommand \enquote  [1]{``#1''}%
\providecommand \bibnamefont  [1]{#1}%
\providecommand \bibfnamefont [1]{#1}%
\providecommand \citenamefont [1]{#1}%
\providecommand \href@noop [0]{\@secondoftwo}%
\providecommand \href [0]{\begingroup \@sanitize@url \@href}%
\providecommand \@href[1]{\@@startlink{#1}\@@href}%
\providecommand \@@href[1]{\endgroup#1\@@endlink}%
\providecommand \@sanitize@url [0]{\catcode `\\12\catcode `\$12\catcode `\&12\catcode `\#12\catcode `\^12\catcode `\_12\catcode `\%12\relax}%
\providecommand \@@startlink[1]{}%
\providecommand \@@endlink[0]{}%
\providecommand \url  [0]{\begingroup\@sanitize@url \@url }%
\providecommand \@url [1]{\endgroup\@href {#1}{\urlprefix }}%
\providecommand \urlprefix  [0]{URL }%
\providecommand \Eprint [0]{\href }%
\providecommand \doibase [0]{http://dx.doi.org/}%
\providecommand \selectlanguage [0]{\@gobble}%
\providecommand \bibinfo  [0]{\@secondoftwo}%
\providecommand \bibfield  [0]{\@secondoftwo}%
\providecommand \translation [1]{[#1]}%
\providecommand \BibitemOpen [0]{}%
\providecommand \bibitemStop [0]{}%
\providecommand \bibitemNoStop [0]{.\EOS\space}%
\providecommand \EOS [0]{\spacefactor3000\relax}%
\providecommand \BibitemShut  [1]{\csname bibitem#1\endcsname}%
\let\auto@bib@innerbib\@empty
\bibitem [{\citenamefont {Du}\ \emph {et~al.}(2016)\citenamefont {Du}, \citenamefont {Maassen}, \citenamefont {Wu}, \citenamefont {Luo}, \citenamefont {Xu},\ and\ \citenamefont {Ye}}]{Du_Nano_Letters_2016}%
  \BibitemOpen
  \bibfield  {author} {\bibinfo {author} {\bibfnamefont {Yuchen}\ \bibnamefont {Du}}, \bibinfo {author} {\bibfnamefont {Jesse}\ \bibnamefont {Maassen}}, \bibinfo {author} {\bibfnamefont {Wangran}\ \bibnamefont {Wu}}, \bibinfo {author} {\bibfnamefont {Zhe}\ \bibnamefont {Luo}}, \bibinfo {author} {\bibfnamefont {Xianfan}\ \bibnamefont {Xu}}, \ and\ \bibinfo {author} {\bibfnamefont {Peide~D.}\ \bibnamefont {Ye}},\ }\bibfield  {title} {\enquote {\bibinfo {title} {Auxetic black phosphorus: A {2D} material with negative poisson's ratio},}\ }\href {\doibase 10.1021/acs.nanolett.6b03607} {\bibfield  {journal} {\bibinfo  {journal} {Nano Letters}\ }\textbf {\bibinfo {volume} {16}},\ \bibinfo {pages} {6701--6708} (\bibinfo {year} {2016})}\BibitemShut {NoStop}%
\bibitem [{\citenamefont {Du}\ \emph {et~al.}(2010)\citenamefont {Du}, \citenamefont {Ouyang}, \citenamefont {Shi},\ and\ \citenamefont {Lei}}]{Du_journal_applied_phys_2010}%
  \BibitemOpen
  \bibfield  {author} {\bibinfo {author} {\bibfnamefont {Yanlan}\ \bibnamefont {Du}}, \bibinfo {author} {\bibfnamefont {Chuying}\ \bibnamefont {Ouyang}}, \bibinfo {author} {\bibfnamefont {Siqi}\ \bibnamefont {Shi}}, \ and\ \bibinfo {author} {\bibfnamefont {Minsheng}\ \bibnamefont {Lei}},\ }\bibfield  {title} {\enquote {\bibinfo {title} {Ab initio studies on atomic and electronic structures of black phosphorus},}\ }\href {\doibase 10.1063/1.3386509} {\bibfield  {journal} {\bibinfo  {journal} {Journal of Applied Physics}\ }\textbf {\bibinfo {volume} {107}},\ \bibinfo {pages} {093718} (\bibinfo {year} {2010})}\BibitemShut {NoStop}%
\bibitem [{\citenamefont {Mu}\ \emph {et~al.}(2019)\citenamefont {Mu}, \citenamefont {Wang},\ and\ \citenamefont {Sun}}]{Mu_Mat_today_phy_2019}%
  \BibitemOpen
  \bibfield  {author} {\bibinfo {author} {\bibfnamefont {X.}~\bibnamefont {Mu}}, \bibinfo {author} {\bibfnamefont {J.}~\bibnamefont {Wang}}, \ and\ \bibinfo {author} {\bibfnamefont {M.}~\bibnamefont {Sun}},\ }\bibfield  {title} {\enquote {\bibinfo {title} {Two-dimensional black phosphorus: physical properties and applications},}\ }\href {\doibase https://doi.org/10.1016/j.mtphys.2019.02.003} {\bibfield  {journal} {\bibinfo  {journal} {Materials Today Physics}\ }\textbf {\bibinfo {volume} {8}},\ \bibinfo {pages} {92--111} (\bibinfo {year} {2019})}\BibitemShut {NoStop}%
\bibitem [{\citenamefont {Appalakondaiah}\ \emph {et~al.}(2012)\citenamefont {Appalakondaiah}, \citenamefont {Vaitheeswaran}, \citenamefont {Leb\`egue}, \citenamefont {Christensen},\ and\ \citenamefont {Svane}}]{Appalakondaiah_PhysRevB_2012}%
  \BibitemOpen
  \bibfield  {author} {\bibinfo {author} {\bibfnamefont {S.}~\bibnamefont {Appalakondaiah}}, \bibinfo {author} {\bibfnamefont {G.}~\bibnamefont {Vaitheeswaran}}, \bibinfo {author} {\bibfnamefont {S.}~\bibnamefont {Leb\`egue}}, \bibinfo {author} {\bibfnamefont {N.~E.}\ \bibnamefont {Christensen}}, \ and\ \bibinfo {author} {\bibfnamefont {A.}~\bibnamefont {Svane}},\ }\bibfield  {title} {\enquote {\bibinfo {title} {Effect of van der {Waals} interactions on the structural and elastic properties of black phosphorus},}\ }\href {\doibase 10.1103/PhysRevB.86.035105} {\bibfield  {journal} {\bibinfo  {journal} {Phys. Rev. B}\ }\textbf {\bibinfo {volume} {86}},\ \bibinfo {pages} {035105} (\bibinfo {year} {2012})}\BibitemShut {NoStop}%
\bibitem [{\citenamefont {Li}\ \emph {et~al.}(2014)\citenamefont {Li}, \citenamefont {Yu}, \citenamefont {Ye}, \citenamefont {Ge}, \citenamefont {Ou}, \citenamefont {Wu}, \citenamefont {Feng}, \citenamefont {Chen},\ and\ \citenamefont {Zhang}}]{Li_Nature_nanotech_2014}%
  \BibitemOpen
  \bibfield  {author} {\bibinfo {author} {\bibfnamefont {Likai}\ \bibnamefont {Li}}, \bibinfo {author} {\bibfnamefont {Yijun}\ \bibnamefont {Yu}}, \bibinfo {author} {\bibfnamefont {Guo~Jun}\ \bibnamefont {Ye}}, \bibinfo {author} {\bibfnamefont {Qingqin}\ \bibnamefont {Ge}}, \bibinfo {author} {\bibfnamefont {Xuedong}\ \bibnamefont {Ou}}, \bibinfo {author} {\bibfnamefont {Hua}\ \bibnamefont {Wu}}, \bibinfo {author} {\bibfnamefont {Donglai}\ \bibnamefont {Feng}}, \bibinfo {author} {\bibfnamefont {Xian~Hui}\ \bibnamefont {Chen}}, \ and\ \bibinfo {author} {\bibfnamefont {Yuanbo}\ \bibnamefont {Zhang}},\ }\bibfield  {title} {\enquote {\bibinfo {title} {Black phosphorus field-effect transistors},}\ }\href {\doibase 10.1038/nnano.2014.35} {\bibfield  {journal} {\bibinfo  {journal} {Nature Nanotechnology}\ }\textbf {\bibinfo {volume} {9}},\ \bibinfo {pages} {372--377} (\bibinfo {year} {2014})}\BibitemShut {NoStop}%
\bibitem [{\citenamefont {Tran}\ \emph {et~al.}(2014)\citenamefont {Tran}, \citenamefont {Soklaski}, \citenamefont {Liang},\ and\ \citenamefont {Yang}}]{Tran_PhysRevB_2014}%
  \BibitemOpen
  \bibfield  {author} {\bibinfo {author} {\bibfnamefont {Vy}~\bibnamefont {Tran}}, \bibinfo {author} {\bibfnamefont {Ryan}\ \bibnamefont {Soklaski}}, \bibinfo {author} {\bibfnamefont {Yufeng}\ \bibnamefont {Liang}}, \ and\ \bibinfo {author} {\bibfnamefont {Li}~\bibnamefont {Yang}},\ }\bibfield  {title} {\enquote {\bibinfo {title} {Layer-controlled band gap and anisotropic excitons in few-layer black phosphorus},}\ }\href {\doibase 10.1103/PhysRevB.89.235319} {\bibfield  {journal} {\bibinfo  {journal} {Phys. Rev. B}\ }\textbf {\bibinfo {volume} {89}},\ \bibinfo {pages} {235319} (\bibinfo {year} {2014})}\BibitemShut {NoStop}%
\bibitem [{\citenamefont {Huang}\ and\ \citenamefont {Li}(2016)}]{Huang_Applied_Physics_Letters_2016}%
  \BibitemOpen
  \bibfield  {author} {\bibinfo {author} {\bibfnamefont {Le}~\bibnamefont {Huang}}\ and\ \bibinfo {author} {\bibfnamefont {Jingbo}\ \bibnamefont {Li}},\ }\bibfield  {title} {\enquote {\bibinfo {title} {Tunable electronic structure of black phosphorus/blue phosphorus van der {Waals} p-n heterostructure},}\ }\href {\doibase 10.1063/1.4942368} {\bibfield  {journal} {\bibinfo  {journal} {Applied Physics Letters}\ }\textbf {\bibinfo {volume} {108}},\ \bibinfo {pages} {083101} (\bibinfo {year} {2016})}\BibitemShut {NoStop}%
\bibitem [{\citenamefont {Warschauer}(1963)}]{Warschauer_Journal_AP_1963}%
  \BibitemOpen
  \bibfield  {author} {\bibinfo {author} {\bibfnamefont {Douglas}\ \bibnamefont {Warschauer}},\ }\bibfield  {title} {\enquote {\bibinfo {title} {Electrical and optical properties of crystalline black phosphorus},}\ }\href {\doibase 10.1063/1.1729699} {\bibfield  {journal} {\bibinfo  {journal} {Journal of Applied Physics}\ }\textbf {\bibinfo {volume} {34}},\ \bibinfo {pages} {1853--1860} (\bibinfo {year} {1963})}\BibitemShut {NoStop}%
\bibitem [{\citenamefont {Guan}\ \emph {et~al.}(2016)\citenamefont {Guan}, \citenamefont {Song}, \citenamefont {Yang},\ and\ \citenamefont {Tom\'anek}}]{Guan_PhysRevB_2016}%
  \BibitemOpen
  \bibfield  {author} {\bibinfo {author} {\bibfnamefont {Jie}\ \bibnamefont {Guan}}, \bibinfo {author} {\bibfnamefont {Wenshen}\ \bibnamefont {Song}}, \bibinfo {author} {\bibfnamefont {Li}~\bibnamefont {Yang}}, \ and\ \bibinfo {author} {\bibfnamefont {David}\ \bibnamefont {Tom\'anek}},\ }\bibfield  {title} {\enquote {\bibinfo {title} {Strain-controlled fundamental gap and structure of bulk black phosphorus},}\ }\href {\doibase 10.1103/PhysRevB.94.045414} {\bibfield  {journal} {\bibinfo  {journal} {Phys. Rev. B}\ }\textbf {\bibinfo {volume} {94}},\ \bibinfo {pages} {045414} (\bibinfo {year} {2016})}\BibitemShut {NoStop}%
\bibitem [{\citenamefont {Pogna}\ \emph {et~al.}(2022)\citenamefont {Pogna}, \citenamefont {Bosak}, \citenamefont {Chumakova}, \citenamefont {Milman}, \citenamefont {Winkler}, \citenamefont {Viti},\ and\ \citenamefont {Vitiello}}]{Ponaga_PhysRevB_2022}%
  \BibitemOpen
  \bibfield  {author} {\bibinfo {author} {\bibfnamefont {Eva A.~A.}\ \bibnamefont {Pogna}}, \bibinfo {author} {\bibfnamefont {Alexe\"{\i}}\ \bibnamefont {Bosak}}, \bibinfo {author} {\bibfnamefont {Alexandra}\ \bibnamefont {Chumakova}}, \bibinfo {author} {\bibfnamefont {Victor}\ \bibnamefont {Milman}}, \bibinfo {author} {\bibfnamefont {Bj\"orn}\ \bibnamefont {Winkler}}, \bibinfo {author} {\bibfnamefont {Leonardo}\ \bibnamefont {Viti}}, \ and\ \bibinfo {author} {\bibfnamefont {Miriam~S.}\ \bibnamefont {Vitiello}},\ }\bibfield  {title} {\enquote {\bibinfo {title} {Lattice dynamics and elastic properties of black phosphorus},}\ }\href {\doibase 10.1103/PhysRevB.105.184306} {\bibfield  {journal} {\bibinfo  {journal} {Phys. Rev. B}\ }\textbf {\bibinfo {volume} {105}},\ \bibinfo {pages} {184306} (\bibinfo {year} {2022})}\BibitemShut {NoStop}%
\bibitem [{\citenamefont {Yi}\ \emph {et~al.}(2019)\citenamefont {Yi}, \citenamefont {Sun}, \citenamefont {Li}, \citenamefont {Chu},\ and\ \citenamefont {Yu}}]{Yi_small_methods_2019}%
  \BibitemOpen
  \bibfield  {author} {\bibinfo {author} {\bibfnamefont {Ya}~\bibnamefont {Yi}}, \bibinfo {author} {\bibfnamefont {Zhengbo}\ \bibnamefont {Sun}}, \bibinfo {author} {\bibfnamefont {Jia}\ \bibnamefont {Li}}, \bibinfo {author} {\bibfnamefont {Paul~K.}\ \bibnamefont {Chu}}, \ and\ \bibinfo {author} {\bibfnamefont {Xue-Feng}\ \bibnamefont {Yu}},\ }\bibfield  {title} {\enquote {\bibinfo {title} {Optical and optoelectronic properties of black phosphorus and recent photonic and optoelectronic applications},}\ }\href {\doibase https://doi.org/10.1002/smtd.201900165} {\bibfield  {journal} {\bibinfo  {journal} {Small Methods}\ }\textbf {\bibinfo {volume} {3}},\ \bibinfo {pages} {1900165} (\bibinfo {year} {2019})}\BibitemShut {NoStop}%
\bibitem [{\citenamefont {Debnath}\ \emph {et~al.}(2018)\citenamefont {Debnath}, \citenamefont {Park},\ and\ \citenamefont {Song}}]{Debnath_small_methods_2018}%
  \BibitemOpen
  \bibfield  {author} {\bibinfo {author} {\bibfnamefont {Pulak~Chandra}\ \bibnamefont {Debnath}}, \bibinfo {author} {\bibfnamefont {Kichul}\ \bibnamefont {Park}}, \ and\ \bibinfo {author} {\bibfnamefont {Yong-Won}\ \bibnamefont {Song}},\ }\bibfield  {title} {\enquote {\bibinfo {title} {Recent advances in black-phosphorus-based photonics and optoelectronics devices},}\ }\href {\doibase https://doi.org/10.1002/smtd.201700315} {\bibfield  {journal} {\bibinfo  {journal} {Small Methods}\ }\textbf {\bibinfo {volume} {2}},\ \bibinfo {pages} {1700315} (\bibinfo {year} {2018})}\BibitemShut {NoStop}%
\bibitem [{\citenamefont {Xiang}\ \emph {et~al.}(2015)\citenamefont {Xiang}, \citenamefont {Han}, \citenamefont {Wu}, \citenamefont {Zhong}, \citenamefont {Liu}, \citenamefont {Lin}, \citenamefont {Zhang}, \citenamefont {Ping~Hu}, \citenamefont {{\"O}zyilmaz}, \citenamefont {Neto}, \citenamefont {Wee},\ and\ \citenamefont {Chen}}]{Xiang_nature_comms_2015}%
  \BibitemOpen
  \bibfield  {author} {\bibinfo {author} {\bibfnamefont {Du}~\bibnamefont {Xiang}}, \bibinfo {author} {\bibfnamefont {Cheng}\ \bibnamefont {Han}}, \bibinfo {author} {\bibfnamefont {Jing}\ \bibnamefont {Wu}}, \bibinfo {author} {\bibfnamefont {Shu}\ \bibnamefont {Zhong}}, \bibinfo {author} {\bibfnamefont {Yiyang}\ \bibnamefont {Liu}}, \bibinfo {author} {\bibfnamefont {Jiadan}\ \bibnamefont {Lin}}, \bibinfo {author} {\bibfnamefont {Xue-Ao}\ \bibnamefont {Zhang}}, \bibinfo {author} {\bibfnamefont {Wen}\ \bibnamefont {Ping~Hu}}, \bibinfo {author} {\bibfnamefont {Barbaros}\ \bibnamefont {{\"O}zyilmaz}}, \bibinfo {author} {\bibfnamefont {A.~H.~Castro}\ \bibnamefont {Neto}}, \bibinfo {author} {\bibfnamefont {Andrew Thye~Shen}\ \bibnamefont {Wee}}, \ and\ \bibinfo {author} {\bibfnamefont {Wei}\ \bibnamefont {Chen}},\ }\bibfield  {title} {\enquote {\bibinfo {title} {Surface transfer doping induced effective modulation on ambipolar characteristics of few-layer black phosphorus},}\ }\href {\doibase 10.1038/ncomms7485}
  {\bibfield  {journal} {\bibinfo  {journal} {Nature Communications}\ }\textbf {\bibinfo {volume} {6}},\ \bibinfo {pages} {6485} (\bibinfo {year} {2015})}\BibitemShut {NoStop}%
\bibitem [{\citenamefont {Koenig}\ \emph {et~al.}(2016)\citenamefont {Koenig}, \citenamefont {Doganov}, \citenamefont {Seixas}, \citenamefont {Carvalho}, \citenamefont {Tan}, \citenamefont {Watanabe}, \citenamefont {Taniguchi}, \citenamefont {Yakovlev}, \citenamefont {Castro~Neto},\ and\ \citenamefont {{\"O}zyilmaz}}]{Koenig_nano_letters_2016}%
  \BibitemOpen
  \bibfield  {author} {\bibinfo {author} {\bibfnamefont {Steven~P.}\ \bibnamefont {Koenig}}, \bibinfo {author} {\bibfnamefont {Rostislav~A.}\ \bibnamefont {Doganov}}, \bibinfo {author} {\bibfnamefont {Leandro}\ \bibnamefont {Seixas}}, \bibinfo {author} {\bibfnamefont {Alexandra}\ \bibnamefont {Carvalho}}, \bibinfo {author} {\bibfnamefont {Jun~You}\ \bibnamefont {Tan}}, \bibinfo {author} {\bibfnamefont {Kenji}\ \bibnamefont {Watanabe}}, \bibinfo {author} {\bibfnamefont {Takashi}\ \bibnamefont {Taniguchi}}, \bibinfo {author} {\bibfnamefont {Nikolai}\ \bibnamefont {Yakovlev}}, \bibinfo {author} {\bibfnamefont {Antonio~H.}\ \bibnamefont {Castro~Neto}}, \ and\ \bibinfo {author} {\bibfnamefont {Barbaros}\ \bibnamefont {{\"O}zyilmaz}},\ }\bibfield  {title} {\enquote {\bibinfo {title} {Electron doping of ultrathin black phosphorus with {Cu} adatoms},}\ }\href {\doibase 10.1021/acs.nanolett.5b03278} {\bibfield  {journal} {\bibinfo  {journal} {Nano Letters}\ }\textbf {\bibinfo {volume} {16}},\ \bibinfo {pages}
  {2145--2151} (\bibinfo {year} {2016})}\BibitemShut {NoStop}%
\bibitem [{\citenamefont {Xu}\ \emph {et~al.}(2016)\citenamefont {Xu}, \citenamefont {Yuan}, \citenamefont {Fei}, \citenamefont {Wang}, \citenamefont {Bao}, \citenamefont {Wang}, \citenamefont {Zhang},\ and\ \citenamefont {Zhang}}]{Xu_small_2016}%
  \BibitemOpen
  \bibfield  {author} {\bibinfo {author} {\bibfnamefont {Yijun}\ \bibnamefont {Xu}}, \bibinfo {author} {\bibfnamefont {Jian}\ \bibnamefont {Yuan}}, \bibinfo {author} {\bibfnamefont {Linfeng}\ \bibnamefont {Fei}}, \bibinfo {author} {\bibfnamefont {Xinliang}\ \bibnamefont {Wang}}, \bibinfo {author} {\bibfnamefont {Qiaoliang}\ \bibnamefont {Bao}}, \bibinfo {author} {\bibfnamefont {Yu}~\bibnamefont {Wang}}, \bibinfo {author} {\bibfnamefont {Kai}\ \bibnamefont {Zhang}}, \ and\ \bibinfo {author} {\bibfnamefont {Yuegang}\ \bibnamefont {Zhang}},\ }\bibfield  {title} {\enquote {\bibinfo {title} {Selenium-doped black phosphorus for high-responsivity {2D} photodetectors},}\ }\href {\doibase https://doi.org/10.1002/smll.201600692} {\bibfield  {journal} {\bibinfo  {journal} {Small}\ }\textbf {\bibinfo {volume} {12}},\ \bibinfo {pages} {5000--5007} (\bibinfo {year} {2016})}\BibitemShut {NoStop}%
\bibitem [{\citenamefont {Chen}\ \emph {et~al.}(2017)\citenamefont {Chen}, \citenamefont {Zhang}, \citenamefont {Cai}, \citenamefont {Jiang}, \citenamefont {Zheng}, \citenamefont {Zhao},\ and\ \citenamefont {Wei}}]{CHEN_carbon_2017}%
  \BibitemOpen
  \bibfield  {author} {\bibinfo {author} {\bibfnamefont {Yang}\ \bibnamefont {Chen}}, \bibinfo {author} {\bibfnamefont {Yingyan}\ \bibnamefont {Zhang}}, \bibinfo {author} {\bibfnamefont {Kun}\ \bibnamefont {Cai}}, \bibinfo {author} {\bibfnamefont {Jinwu}\ \bibnamefont {Jiang}}, \bibinfo {author} {\bibfnamefont {Jin-Cheng}\ \bibnamefont {Zheng}}, \bibinfo {author} {\bibfnamefont {Junhua}\ \bibnamefont {Zhao}}, \ and\ \bibinfo {author} {\bibfnamefont {Ning}\ \bibnamefont {Wei}},\ }\bibfield  {title} {\enquote {\bibinfo {title} {Interfacial thermal conductance in graphene/black phosphorus heterogeneous structures},}\ }\href {\doibase https://doi.org/10.1016/j.carbon.2017.03.011} {\bibfield  {journal} {\bibinfo  {journal} {Carbon}\ }\textbf {\bibinfo {volume} {117}},\ \bibinfo {pages} {399--410} (\bibinfo {year} {2017})}\BibitemShut {NoStop}%
\bibitem [{\citenamefont {Wu}\ \emph {et~al.}(2021)\citenamefont {Wu}, \citenamefont {Ma}, \citenamefont {He}, \citenamefont {Wu}, \citenamefont {Tang}, \citenamefont {Yu}, \citenamefont {Wu}, \citenamefont {Chen},\ and\ \citenamefont {Bao}}]{Wu_Angewandte_chemie_international_2021}%
  \BibitemOpen
  \bibfield  {author} {\bibinfo {author} {\bibfnamefont {Tianyu}\ \bibnamefont {Wu}}, \bibinfo {author} {\bibfnamefont {Ziyang}\ \bibnamefont {Ma}}, \bibinfo {author} {\bibfnamefont {Yunya}\ \bibnamefont {He}}, \bibinfo {author} {\bibfnamefont {Xingjiang}\ \bibnamefont {Wu}}, \bibinfo {author} {\bibfnamefont {Bao}\ \bibnamefont {Tang}}, \bibinfo {author} {\bibfnamefont {Ziyi}\ \bibnamefont {Yu}}, \bibinfo {author} {\bibfnamefont {Guan}\ \bibnamefont {Wu}}, \bibinfo {author} {\bibfnamefont {Su}~\bibnamefont {Chen}}, \ and\ \bibinfo {author} {\bibfnamefont {Ningzhong}\ \bibnamefont {Bao}},\ }\bibfield  {title} {\enquote {\bibinfo {title} {A covalent black phosphorus/metal–organic framework hetero-nanostructure for high-performance flexible supercapacitors},}\ }\href {\doibase https://doi.org/10.1002/anie.202101648} {\bibfield  {journal} {\bibinfo  {journal} {Angewandte Chemie International Edition}\ }\textbf {\bibinfo {volume} {60}},\ \bibinfo {pages} {10366--10374} (\bibinfo {year} {2021})}\BibitemShut
  {NoStop}%
\bibitem [{\citenamefont {Li}\ \emph {et~al.}(2018)\citenamefont {Li}, \citenamefont {Sun}, \citenamefont {Shahi}, \citenamefont {Gao}, \citenamefont {MacDonald}, \citenamefont {Uwatoko}, \citenamefont {Xiang}, \citenamefont {Goodenough}, \citenamefont {Cheng},\ and\ \citenamefont {Zhou}}]{Li_national_academy_sciences_2018}%
  \BibitemOpen
  \bibfield  {author} {\bibinfo {author} {\bibfnamefont {Xiang}\ \bibnamefont {Li}}, \bibinfo {author} {\bibfnamefont {Jianping}\ \bibnamefont {Sun}}, \bibinfo {author} {\bibfnamefont {Prashant}\ \bibnamefont {Shahi}}, \bibinfo {author} {\bibfnamefont {Miao}\ \bibnamefont {Gao}}, \bibinfo {author} {\bibfnamefont {Allan~H.}\ \bibnamefont {MacDonald}}, \bibinfo {author} {\bibfnamefont {Yoshiya}\ \bibnamefont {Uwatoko}}, \bibinfo {author} {\bibfnamefont {Tao}\ \bibnamefont {Xiang}}, \bibinfo {author} {\bibfnamefont {John~B.}\ \bibnamefont {Goodenough}}, \bibinfo {author} {\bibfnamefont {Jinguang}\ \bibnamefont {Cheng}}, \ and\ \bibinfo {author} {\bibfnamefont {Jianshi}\ \bibnamefont {Zhou}},\ }\bibfield  {title} {\enquote {\bibinfo {title} {Pressure-induced phase transitions and superconductivity in a black phosphorus single crystal},}\ }\href {\doibase 10.1073/pnas.1810726115} {\bibfield  {journal} {\bibinfo  {journal} {Proceedings of the National Academy of Sciences}\ }\textbf {\bibinfo {volume} {115}},\
  \bibinfo {pages} {9935--9940} (\bibinfo {year} {2018})}\BibitemShut {NoStop}%
\bibitem [{\citenamefont {He}\ \emph {et~al.}(2012)\citenamefont {He}, \citenamefont {Sun}, \citenamefont {Liu},\ and\ \citenamefont {Zhang}}]{he_high_k_gate_dielectrics_2012}%
  \BibitemOpen
  \bibfield  {author} {\bibinfo {author} {\bibfnamefont {Gang}\ \bibnamefont {He}}, \bibinfo {author} {\bibfnamefont {Zhaoqi}\ \bibnamefont {Sun}}, \bibinfo {author} {\bibfnamefont {Mao}\ \bibnamefont {Liu}}, \ and\ \bibinfo {author} {\bibfnamefont {Lide}\ \bibnamefont {Zhang}},\ }\bibfield  {title} {\enquote {\bibinfo {title} {Scaling and limitation of {Si}-based {C}{M}{O}{S}},}\ }\href@noop {} {\bibfield  {journal} {\bibinfo  {journal} {High-k Gate Dielectrics for CMOS Technology}\ ,\ \bibinfo {pages} {1--29}} (\bibinfo {year} {2012})}\BibitemShut {NoStop}%
\bibitem [{\citenamefont {van Druenen}(2020)}]{vanDruenen_AdvMatInterfaces_2020}%
  \BibitemOpen
  \bibfield  {author} {\bibinfo {author} {\bibfnamefont {M.}~\bibnamefont {van Druenen}},\ }\bibfield  {title} {\enquote {\bibinfo {title} {Degradation of black phosphorus and strategies to enhance its ambient lifetime},}\ }\href {\doibase https://doi.org/10.1002/admi.202001102} {\bibfield  {journal} {\bibinfo  {journal} {Advanced Materials Interfaces}\ }\textbf {\bibinfo {volume} {7}},\ \bibinfo {pages} {2001102} (\bibinfo {year} {2020})}\BibitemShut {NoStop}%
\bibitem [{\citenamefont {Island}\ \emph {et~al.}(2015)\citenamefont {Island}, \citenamefont {Steele}, \citenamefont {Zant},\ and\ \citenamefont {Castellanos-Gomez}}]{Island_2D_materials_2015}%
  \BibitemOpen
  \bibfield  {author} {\bibinfo {author} {\bibfnamefont {Joshua~O}\ \bibnamefont {Island}}, \bibinfo {author} {\bibfnamefont {Gary~A}\ \bibnamefont {Steele}}, \bibinfo {author} {\bibfnamefont {Herre S J van~der}\ \bibnamefont {Zant}}, \ and\ \bibinfo {author} {\bibfnamefont {Andres}\ \bibnamefont {Castellanos-Gomez}},\ }\bibfield  {title} {\enquote {\bibinfo {title} {Environmental instability of few-layer black phosphorus},}\ }\href {\doibase 10.1088/2053-1583/2/1/011002} {\bibfield  {journal} {\bibinfo  {journal} {2D Materials}\ }\textbf {\bibinfo {volume} {2}},\ \bibinfo {pages} {011002} (\bibinfo {year} {2015})}\BibitemShut {NoStop}%
\bibitem [{\citenamefont {Lv}\ \emph {et~al.}(2018)\citenamefont {Lv}, \citenamefont {Yang}, \citenamefont {Wang}, \citenamefont {Wan}, \citenamefont {Ge}, \citenamefont {Yang}, \citenamefont {Hao}, \citenamefont {Xiang}, \citenamefont {Zhang}, \citenamefont {Zeng},\ and\ \citenamefont {Liu}}]{Lv_ACS_applied_Materials_2018}%
  \BibitemOpen
  \bibfield  {author} {\bibinfo {author} {\bibfnamefont {Weiming}\ \bibnamefont {Lv}}, \bibinfo {author} {\bibfnamefont {Bingchao}\ \bibnamefont {Yang}}, \bibinfo {author} {\bibfnamefont {Bochong}\ \bibnamefont {Wang}}, \bibinfo {author} {\bibfnamefont {Wenhui}\ \bibnamefont {Wan}}, \bibinfo {author} {\bibfnamefont {Yanfeng}\ \bibnamefont {Ge}}, \bibinfo {author} {\bibfnamefont {Ruilong}\ \bibnamefont {Yang}}, \bibinfo {author} {\bibfnamefont {Chunxue}\ \bibnamefont {Hao}}, \bibinfo {author} {\bibfnamefont {Jianyong}\ \bibnamefont {Xiang}}, \bibinfo {author} {\bibfnamefont {Baoshun}\ \bibnamefont {Zhang}}, \bibinfo {author} {\bibfnamefont {Zhongming}\ \bibnamefont {Zeng}}, \ and\ \bibinfo {author} {\bibfnamefont {Zhongyuan}\ \bibnamefont {Liu}},\ }\bibfield  {title} {\enquote {\bibinfo {title} {Sulfur-doped black phosphorus field-effect transistors with enhanced stability},}\ }\href {\doibase 10.1021/acsami.7b19169} {\bibfield  {journal} {\bibinfo  {journal} {ACS Applied Materials {\&} Interfaces}\ }\textbf
  {\bibinfo {volume} {10}},\ \bibinfo {pages} {9663--9668} (\bibinfo {year} {2018})}\BibitemShut {NoStop}%
\bibitem [{\citenamefont {Sarswat}\ \emph {et~al.}(2016)\citenamefont {Sarswat}, \citenamefont {Sarkar}, \citenamefont {Cho}, \citenamefont {Bhattacharyya},\ and\ \citenamefont {Free}}]{Sarswat_electrochem_society_2016}%
  \BibitemOpen
  \bibfield  {author} {\bibinfo {author} {\bibfnamefont {Prashant~K}\ \bibnamefont {Sarswat}}, \bibinfo {author} {\bibfnamefont {Sayan}\ \bibnamefont {Sarkar}}, \bibinfo {author} {\bibfnamefont {Jaehun}\ \bibnamefont {Cho}}, \bibinfo {author} {\bibfnamefont {Dhiman}\ \bibnamefont {Bhattacharyya}}, \ and\ \bibinfo {author} {\bibfnamefont {Michael~L.}\ \bibnamefont {Free}},\ }\bibfield  {title} {\enquote {\bibinfo {title} {Structural and electrical irregularities caused by selected dopants in black-phosphorus},}\ }\href {\doibase 10.1149/2.0061611jss} {\bibfield  {journal} {\bibinfo  {journal} {ECS Journal of Solid State Science and Technology}\ }\textbf {\bibinfo {volume} {5}},\ \bibinfo {pages} {Q3026} (\bibinfo {year} {2016})}\BibitemShut {NoStop}%
\bibitem [{\citenamefont {Yang}\ \emph {et~al.}(2016)\citenamefont {Yang}, \citenamefont {Wan}, \citenamefont {Zhou}, \citenamefont {Wang}, \citenamefont {Hu}, \citenamefont {Lv}, \citenamefont {Chen}, \citenamefont {Zeng}, \citenamefont {Wen}, \citenamefont {Xiang}, \citenamefont {Yuan}, \citenamefont {Wang}, \citenamefont {Zhang}, \citenamefont {Wang}, \citenamefont {Zhang}, \citenamefont {Xu}, \citenamefont {Zhao}, \citenamefont {Tian},\ and\ \citenamefont {Liu}}]{Yang_advanced_materials_2016}%
  \BibitemOpen
  \bibfield  {author} {\bibinfo {author} {\bibfnamefont {Bingchao}\ \bibnamefont {Yang}}, \bibinfo {author} {\bibfnamefont {Bensong}\ \bibnamefont {Wan}}, \bibinfo {author} {\bibfnamefont {Qionghua}\ \bibnamefont {Zhou}}, \bibinfo {author} {\bibfnamefont {Yue}\ \bibnamefont {Wang}}, \bibinfo {author} {\bibfnamefont {Wentao}\ \bibnamefont {Hu}}, \bibinfo {author} {\bibfnamefont {Weiming}\ \bibnamefont {Lv}}, \bibinfo {author} {\bibfnamefont {Qian}\ \bibnamefont {Chen}}, \bibinfo {author} {\bibfnamefont {Zhongming}\ \bibnamefont {Zeng}}, \bibinfo {author} {\bibfnamefont {Fusheng}\ \bibnamefont {Wen}}, \bibinfo {author} {\bibfnamefont {Jianyong}\ \bibnamefont {Xiang}}, \bibinfo {author} {\bibfnamefont {Shijun}\ \bibnamefont {Yuan}}, \bibinfo {author} {\bibfnamefont {Jinlan}\ \bibnamefont {Wang}}, \bibinfo {author} {\bibfnamefont {Baoshun}\ \bibnamefont {Zhang}}, \bibinfo {author} {\bibfnamefont {Wenhong}\ \bibnamefont {Wang}}, \bibinfo {author} {\bibfnamefont {Junying}\ \bibnamefont {Zhang}}, \bibinfo {author}
  {\bibfnamefont {Bo}~\bibnamefont {Xu}}, \bibinfo {author} {\bibfnamefont {Zhisheng}\ \bibnamefont {Zhao}}, \bibinfo {author} {\bibfnamefont {Yongjun}\ \bibnamefont {Tian}}, \ and\ \bibinfo {author} {\bibfnamefont {Zhongyuan}\ \bibnamefont {Liu}},\ }\bibfield  {title} {\enquote {\bibinfo {title} {Te-doped black phosphorus field-effect transistors},}\ }\href {\doibase https://doi.org/10.1002/adma.201603723} {\bibfield  {journal} {\bibinfo  {journal} {Advanced Materials}\ }\textbf {\bibinfo {volume} {28}},\ \bibinfo {pages} {9408--9415} (\bibinfo {year} {2016})}\BibitemShut {NoStop}%
\bibitem [{\citenamefont {Doganov}\ \emph {et~al.}(2015)\citenamefont {Doganov}, \citenamefont {O'Farrell}, \citenamefont {Koenig}, \citenamefont {Yeo}, \citenamefont {Ziletti}, \citenamefont {Carvalho}, \citenamefont {Campbell}, \citenamefont {Coker}, \citenamefont {Watanabe}, \citenamefont {Taniguchi}, \citenamefont {Neto},\ and\ \citenamefont {{\"O}zyilmaz}}]{Doganov_nature_comms_2015}%
  \BibitemOpen
  \bibfield  {author} {\bibinfo {author} {\bibfnamefont {Rostislav~A.}\ \bibnamefont {Doganov}}, \bibinfo {author} {\bibfnamefont {Eoin C.~T.}\ \bibnamefont {O'Farrell}}, \bibinfo {author} {\bibfnamefont {Steven~P.}\ \bibnamefont {Koenig}}, \bibinfo {author} {\bibfnamefont {Yuting}\ \bibnamefont {Yeo}}, \bibinfo {author} {\bibfnamefont {Angelo}\ \bibnamefont {Ziletti}}, \bibinfo {author} {\bibfnamefont {Alexandra}\ \bibnamefont {Carvalho}}, \bibinfo {author} {\bibfnamefont {David~K.}\ \bibnamefont {Campbell}}, \bibinfo {author} {\bibfnamefont {David~F.}\ \bibnamefont {Coker}}, \bibinfo {author} {\bibfnamefont {Kenji}\ \bibnamefont {Watanabe}}, \bibinfo {author} {\bibfnamefont {Takashi}\ \bibnamefont {Taniguchi}}, \bibinfo {author} {\bibfnamefont {Antonio H.~Castro}\ \bibnamefont {Neto}}, \ and\ \bibinfo {author} {\bibfnamefont {Barbaros}\ \bibnamefont {{\"O}zyilmaz}},\ }\bibfield  {title} {\enquote {\bibinfo {title} {Transport properties of pristine few-layer black phosphorus by van der {Waals} passivation in
  an inert atmosphere},}\ }\href {\doibase 10.1038/ncomms7647} {\bibfield  {journal} {\bibinfo  {journal} {Nature Communications}\ }\textbf {\bibinfo {volume} {6}},\ \bibinfo {pages} {6647} (\bibinfo {year} {2015})}\BibitemShut {NoStop}%
\bibitem [{\citenamefont {Wan}\ \emph {et~al.}(2015)\citenamefont {Wan}, \citenamefont {Yang}, \citenamefont {Wang}, \citenamefont {Zhang}, \citenamefont {Zeng}, \citenamefont {Liu},\ and\ \citenamefont {Wang}}]{Wan_nanotechnology_2015}%
  \BibitemOpen
  \bibfield  {author} {\bibinfo {author} {\bibfnamefont {Bensong}\ \bibnamefont {Wan}}, \bibinfo {author} {\bibfnamefont {Bingchao}\ \bibnamefont {Yang}}, \bibinfo {author} {\bibfnamefont {Yue}\ \bibnamefont {Wang}}, \bibinfo {author} {\bibfnamefont {Junying}\ \bibnamefont {Zhang}}, \bibinfo {author} {\bibfnamefont {Zhongming}\ \bibnamefont {Zeng}}, \bibinfo {author} {\bibfnamefont {Zhongyuan}\ \bibnamefont {Liu}}, \ and\ \bibinfo {author} {\bibfnamefont {Wenhong}\ \bibnamefont {Wang}},\ }\bibfield  {title} {\enquote {\bibinfo {title} {Enhanced stability of black phosphorus field-effect transistors with {Si}{O}$_2$ passivation},}\ }\href {\doibase 10.1088/0957-4484/26/43/435702} {\bibfield  {journal} {\bibinfo  {journal} {Nanotechnology}\ }\textbf {\bibinfo {volume} {26}},\ \bibinfo {pages} {435702} (\bibinfo {year} {2015})}\BibitemShut {NoStop}%
\bibitem [{\citenamefont {Song}\ \emph {et~al.}(2019)\citenamefont {Song}, \citenamefont {Chen}, \citenamefont {Li}, \citenamefont {Xu}, \citenamefont {Liu}, \citenamefont {Wang},\ and\ \citenamefont {Xia}}]{song_Adv_functional_materials_2019}%
  \BibitemOpen
  \bibfield  {author} {\bibinfo {author} {\bibfnamefont {Tianbing}\ \bibnamefont {Song}}, \bibinfo {author} {\bibfnamefont {Hai}\ \bibnamefont {Chen}}, \bibinfo {author} {\bibfnamefont {Zhi}\ \bibnamefont {Li}}, \bibinfo {author} {\bibfnamefont {Qunjie}\ \bibnamefont {Xu}}, \bibinfo {author} {\bibfnamefont {Haimei}\ \bibnamefont {Liu}}, \bibinfo {author} {\bibfnamefont {Yonggang}\ \bibnamefont {Wang}}, \ and\ \bibinfo {author} {\bibfnamefont {Yongyao}\ \bibnamefont {Xia}},\ }\bibfield  {title} {\enquote {\bibinfo {title} {Creating an air-stable sulfur-doped black phosphorus-tio2 composite as high-performance anode material for sodium-ion storage},}\ }\href@noop {} {\bibfield  {journal} {\bibinfo  {journal} {Advanced Functional Materials}\ }\textbf {\bibinfo {volume} {29}},\ \bibinfo {pages} {1900535} (\bibinfo {year} {2019})}\BibitemShut {NoStop}%
\bibitem [{\citenamefont {Cartz}\ \emph {et~al.}(1979)\citenamefont {Cartz}, \citenamefont {Srinivasa}, \citenamefont {Riedner}, \citenamefont {Jorgensen},\ and\ \citenamefont {Worlton}}]{Cartz_Journal_Chemical_Physics_1979}%
  \BibitemOpen
  \bibfield  {author} {\bibinfo {author} {\bibfnamefont {L.}~\bibnamefont {Cartz}}, \bibinfo {author} {\bibfnamefont {S.~R.}\ \bibnamefont {Srinivasa}}, \bibinfo {author} {\bibfnamefont {R.~J.}\ \bibnamefont {Riedner}}, \bibinfo {author} {\bibfnamefont {J.~D.}\ \bibnamefont {Jorgensen}}, \ and\ \bibinfo {author} {\bibfnamefont {T.~G.}\ \bibnamefont {Worlton}},\ }\bibfield  {title} {\enquote {\bibinfo {title} {Effect of pressure on bonding in black phosphorus},}\ }\href {\doibase 10.1063/1.438523} {\bibfield  {journal} {\bibinfo  {journal} {The Journal of Chemical Physics}\ }\textbf {\bibinfo {volume} {71}},\ \bibinfo {pages} {1718--1721} (\bibinfo {year} {1979})}\BibitemShut {NoStop}%
\bibitem [{\citenamefont {Kresse}\ and\ \citenamefont {Furthm\"uller}(1996{\natexlab{a}})}]{Kresse96a}%
  \BibitemOpen
  \bibfield  {author} {\bibinfo {author} {\bibfnamefont {G.}~\bibnamefont {Kresse}}\ and\ \bibinfo {author} {\bibfnamefont {J.}~\bibnamefont {Furthm\"uller}},\ }\bibfield  {title} {\enquote {\bibinfo {title} {Efficient iterative schemes for {$ab$ $initio$} total-energy calculations using a plane-wave basis set},}\ }\href@noop {} {\bibfield  {journal} {\bibinfo  {journal} {Phys. Rev. B}\ }\textbf {\bibinfo {volume} {54}},\ \bibinfo {pages} {11169--11186} (\bibinfo {year} {1996}{\natexlab{a}})}\BibitemShut {NoStop}%
\bibitem [{\citenamefont {Kresse}\ and\ \citenamefont {Furthm\"uller}(1996{\natexlab{b}})}]{Kresse96b}%
  \BibitemOpen
  \bibfield  {author} {\bibinfo {author} {\bibfnamefont {G.}~\bibnamefont {Kresse}}\ and\ \bibinfo {author} {\bibfnamefont {J.}~\bibnamefont {Furthm\"uller}},\ }\bibfield  {title} {\enquote {\bibinfo {title} {Efficiency of {$ab$-$initio$} total energy calculations for metals and semiconductors using a plane-wave basis set},}\ }\href@noop {} {\bibfield  {journal} {\bibinfo  {journal} {Comput. Mater. Sci.}\ }\textbf {\bibinfo {volume} {6}},\ \bibinfo {pages} {15 -- 50} (\bibinfo {year} {1996}{\natexlab{b}})}\BibitemShut {NoStop}%
\bibitem [{\citenamefont {Kresse}\ and\ \citenamefont {Joubert}(1999)}]{KressePAW}%
  \BibitemOpen
  \bibfield  {author} {\bibinfo {author} {\bibfnamefont {G.}~\bibnamefont {Kresse}}\ and\ \bibinfo {author} {\bibfnamefont {D.}~\bibnamefont {Joubert}},\ }\bibfield  {title} {\enquote {\bibinfo {title} {From ultrasoft pseudopotentials to the projector augmented-wave method},}\ }\href@noop {} {\bibfield  {journal} {\bibinfo  {journal} {Phys. Rev. B}\ }\textbf {\bibinfo {volume} {59}},\ \bibinfo {pages} {1758--1775} (\bibinfo {year} {1999})}\BibitemShut {NoStop}%
\bibitem [{\citenamefont {Bl\"ochl}(1994)}]{Blochl94}%
  \BibitemOpen
  \bibfield  {author} {\bibinfo {author} {\bibfnamefont {P.~E.}\ \bibnamefont {Bl\"ochl}},\ }\bibfield  {title} {\enquote {\bibinfo {title} {Projector augmented-wave method},}\ }\href@noop {} {\bibfield  {journal} {\bibinfo  {journal} {Phys. Rev. B}\ }\textbf {\bibinfo {volume} {50}},\ \bibinfo {pages} {17953--17979} (\bibinfo {year} {1994})}\BibitemShut {NoStop}%
\bibitem [{\citenamefont {Perdew}\ \emph {et~al.}(2008)\citenamefont {Perdew}, \citenamefont {Ruzsinszky}, \citenamefont {Csonka}, \citenamefont {Vydrov}, \citenamefont {Scuseria}, \citenamefont {Constantin}, \citenamefont {Zhou},\ and\ \citenamefont {Burke}}]{PBEsol}%
  \BibitemOpen
  \bibfield  {author} {\bibinfo {author} {\bibfnamefont {John~P.}\ \bibnamefont {Perdew}}, \bibinfo {author} {\bibfnamefont {Adrienn}\ \bibnamefont {Ruzsinszky}}, \bibinfo {author} {\bibfnamefont {G\'abor~I.}\ \bibnamefont {Csonka}}, \bibinfo {author} {\bibfnamefont {Oleg~A.}\ \bibnamefont {Vydrov}}, \bibinfo {author} {\bibfnamefont {Gustavo~E.}\ \bibnamefont {Scuseria}}, \bibinfo {author} {\bibfnamefont {Lucian~A.}\ \bibnamefont {Constantin}}, \bibinfo {author} {\bibfnamefont {Xiaolan}\ \bibnamefont {Zhou}}, \ and\ \bibinfo {author} {\bibfnamefont {Kieron}\ \bibnamefont {Burke}},\ }\bibfield  {title} {\enquote {\bibinfo {title} {Restoring the density-gradient expansion for exchange in solids and surfaces},}\ }\href@noop {} {\bibfield  {journal} {\bibinfo  {journal} {Phys. Rev. Lett.}\ }\textbf {\bibinfo {volume} {100}},\ \bibinfo {pages} {136406} (\bibinfo {year} {2008})}\BibitemShut {NoStop}%
\bibitem [{\citenamefont {Grimme}\ \emph {et~al.}(2010)\citenamefont {Grimme}, \citenamefont {Antony}, \citenamefont {Ehrlich},\ and\ \citenamefont {Krieg}}]{Grimme_DFT_D3_2010}%
  \BibitemOpen
  \bibfield  {author} {\bibinfo {author} {\bibfnamefont {Stefan}\ \bibnamefont {Grimme}}, \bibinfo {author} {\bibfnamefont {Jens}\ \bibnamefont {Antony}}, \bibinfo {author} {\bibfnamefont {Stephan}\ \bibnamefont {Ehrlich}}, \ and\ \bibinfo {author} {\bibfnamefont {Helge}\ \bibnamefont {Krieg}},\ }\bibfield  {title} {\enquote {\bibinfo {title} {A consistent and accurate ab initio parametrization of density functional dispersion correction (dft-d) for the 94 elements h-pu},}\ }\href {\doibase 10.1063/1.3382344} {\bibfield  {journal} {\bibinfo  {journal} {The Journal of Chemical Physics}\ }\textbf {\bibinfo {volume} {132}},\ \bibinfo {pages} {154104} (\bibinfo {year} {2010})}\BibitemShut {NoStop}%
\bibitem [{\citenamefont {Grimme}\ \emph {et~al.}(2011)\citenamefont {Grimme}, \citenamefont {Ehrlich},\ and\ \citenamefont {Goerigk}}]{Grimme_DFT_D3_2011}%
  \BibitemOpen
  \bibfield  {author} {\bibinfo {author} {\bibfnamefont {Stefan}\ \bibnamefont {Grimme}}, \bibinfo {author} {\bibfnamefont {Stephan}\ \bibnamefont {Ehrlich}}, \ and\ \bibinfo {author} {\bibfnamefont {Lars}\ \bibnamefont {Goerigk}},\ }\bibfield  {title} {\enquote {\bibinfo {title} {Effect of the damping function in dispersion corrected density functional theory},}\ }\href {\doibase https://doi.org/10.1002/jcc.21759} {\bibfield  {journal} {\bibinfo  {journal} {Journal of Computational Chemistry}\ }\textbf {\bibinfo {volume} {32}},\ \bibinfo {pages} {1456--1465} (\bibinfo {year} {2011})}\BibitemShut {NoStop}%
\bibitem [{\citenamefont {Singh}\ \emph {et~al.}(2021)\citenamefont {Singh}, \citenamefont {Lang}, \citenamefont {Dovale-Farelo}, \citenamefont {Herath}, \citenamefont {Tavadze}, \citenamefont {Coudert},\ and\ \citenamefont {Romero}}]{mechelastic}%
  \BibitemOpen
  \bibfield  {author} {\bibinfo {author} {\bibfnamefont {Sobhit}\ \bibnamefont {Singh}}, \bibinfo {author} {\bibfnamefont {Logan}\ \bibnamefont {Lang}}, \bibinfo {author} {\bibfnamefont {Viviana}\ \bibnamefont {Dovale-Farelo}}, \bibinfo {author} {\bibfnamefont {Uthpala}\ \bibnamefont {Herath}}, \bibinfo {author} {\bibfnamefont {Pedram}\ \bibnamefont {Tavadze}}, \bibinfo {author} {\bibfnamefont {François-Xavier}\ \bibnamefont {Coudert}}, \ and\ \bibinfo {author} {\bibfnamefont {Aldo~H.}\ \bibnamefont {Romero}},\ }\bibfield  {title} {\enquote {\bibinfo {title} {Mechelastic: A python library for analysis of mechanical and elastic properties of bulk and {2D} materials},}\ }\href {\doibase https://doi.org/10.1016/j.cpc.2021.108068} {\bibfield  {journal} {\bibinfo  {journal} {Computer Physics Communications}\ }\textbf {\bibinfo {volume} {267}},\ \bibinfo {pages} {108068} (\bibinfo {year} {2021})}\BibitemShut {NoStop}%
\bibitem [{\citenamefont {Singh}\ \emph {et~al.}(2018)\citenamefont {Singh}, \citenamefont {Valencia-Jaime}, \citenamefont {Pavlic},\ and\ \citenamefont {Romero}}]{SinghPRB2018_meche}%
  \BibitemOpen
  \bibfield  {author} {\bibinfo {author} {\bibfnamefont {Sobhit}\ \bibnamefont {Singh}}, \bibinfo {author} {\bibfnamefont {Irais}\ \bibnamefont {Valencia-Jaime}}, \bibinfo {author} {\bibfnamefont {Olivia}\ \bibnamefont {Pavlic}}, \ and\ \bibinfo {author} {\bibfnamefont {Aldo~H.}\ \bibnamefont {Romero}},\ }\bibfield  {title} {\enquote {\bibinfo {title} {Elastic, mechanical, and thermodynamic properties of {Bi}-{Sb} binaries: Effect of spin-orbit coupling},}\ }\href {\doibase 10.1103/PhysRevB.97.054108} {\bibfield  {journal} {\bibinfo  {journal} {Phys. Rev. B}\ }\textbf {\bibinfo {volume} {97}},\ \bibinfo {pages} {054108} (\bibinfo {year} {2018})}\BibitemShut {NoStop}%
\bibitem [{\citenamefont {Gaillac}\ \emph {et~al.}(2016)\citenamefont {Gaillac}, \citenamefont {Pullumbi},\ and\ \citenamefont {Coudert}}]{Gaillac_2016}%
  \BibitemOpen
  \bibfield  {author} {\bibinfo {author} {\bibfnamefont {Romain}\ \bibnamefont {Gaillac}}, \bibinfo {author} {\bibfnamefont {Pluton}\ \bibnamefont {Pullumbi}}, \ and\ \bibinfo {author} {\bibfnamefont {François-Xavier}\ \bibnamefont {Coudert}},\ }\bibfield  {title} {\enquote {\bibinfo {title} {Elate: an open-source online application for analysis and visualization of elastic tensors},}\ }\href {\doibase 10.1088/0953-8984/28/27/275201} {\bibfield  {journal} {\bibinfo  {journal} {Journal of Physics: Condensed Matter}\ }\textbf {\bibinfo {volume} {28}},\ \bibinfo {pages} {275201} (\bibinfo {year} {2016})}\BibitemShut {NoStop}%
\bibitem [{\citenamefont {Momma}\ and\ \citenamefont {Izumi}(2011)}]{VESTA}%
  \BibitemOpen
  \bibfield  {author} {\bibinfo {author} {\bibfnamefont {K.}~\bibnamefont {Momma}}\ and\ \bibinfo {author} {\bibfnamefont {F.}~\bibnamefont {Izumi}},\ }\bibfield  {title} {\enquote {\bibinfo {title} {Vesta 3 for three-dimensional visualization of crystal, volumetric and morphology data},}\ }\href@noop {} {\bibfield  {journal} {\bibinfo  {journal} {J. Appl. Crystallogr.}\ }\textbf {\bibinfo {volume} {44}},\ \bibinfo {pages} {1272--1276} (\bibinfo {year} {2011})}\BibitemShut {NoStop}%
\bibitem [{\citenamefont {Heyd}\ \emph {et~al.}(2003)\citenamefont {Heyd}, \citenamefont {Scuseria},\ and\ \citenamefont {Ernzerhof}}]{HSE06_1}%
  \BibitemOpen
  \bibfield  {author} {\bibinfo {author} {\bibfnamefont {Jochen}\ \bibnamefont {Heyd}}, \bibinfo {author} {\bibfnamefont {Gustavo~E.}\ \bibnamefont {Scuseria}}, \ and\ \bibinfo {author} {\bibfnamefont {Matthias}\ \bibnamefont {Ernzerhof}},\ }\bibfield  {title} {\enquote {\bibinfo {title} {{Hybrid functionals based on a screened Coulomb potential}},}\ }\href {\doibase 10.1063/1.1564060} {\bibfield  {journal} {\bibinfo  {journal} {The Journal of Chemical Physics}\ }\textbf {\bibinfo {volume} {118}},\ \bibinfo {pages} {8207--8215} (\bibinfo {year} {2003})}\BibitemShut {NoStop}%
\bibitem [{\citenamefont {Heyd}\ \emph {et~al.}(2006)\citenamefont {Heyd}, \citenamefont {Scuseria},\ and\ \citenamefont {Ernzerhof}}]{HSE06_2}%
  \BibitemOpen
  \bibfield  {author} {\bibinfo {author} {\bibfnamefont {Jochen}\ \bibnamefont {Heyd}}, \bibinfo {author} {\bibfnamefont {Gustavo~E.}\ \bibnamefont {Scuseria}}, \ and\ \bibinfo {author} {\bibfnamefont {Matthias}\ \bibnamefont {Ernzerhof}},\ }\bibfield  {title} {\enquote {\bibinfo {title} {{Erratum: “Hybrid functionals based on a screened Coulomb potential”}},}\ }\href {\doibase 10.1063/1.2204597} {\bibfield  {journal} {\bibinfo  {journal} {The Journal of Chemical Physics}\ }\textbf {\bibinfo {volume} {124}},\ \bibinfo {pages} {219906} (\bibinfo {year} {2006})}\BibitemShut {NoStop}%
\bibitem [{SM()}]{SM}%
  \BibitemOpen
  \href@noop {} {}\bibinfo {note} {See Supplemental Material file at [href] for additional details on the structural configurations, electron localization function plots, and elastic properties of the sulfur-substituted black phosphorus structures with high sulfur content.}\BibitemShut {Stop}%
\bibitem [{\citenamefont {Nye}(1985)}]{nye1985physical}%
  \BibitemOpen
  \bibfield  {author} {\bibinfo {author} {\bibfnamefont {J.F.}\ \bibnamefont {Nye}},\ }\href {https://books.google.com/books?id=ugwql-uVB44C} {\emph {\bibinfo {title} {Physical Properties of Crystals: Their Representation by Tensors and Matrices}}},\ Oxford science publications\ (\bibinfo  {publisher} {Clarendon Press},\ \bibinfo {year} {1985})\BibitemShut {NoStop}%
\bibitem [{\citenamefont {Hill}(1952)}]{hill1952elastic}%
  \BibitemOpen
  \bibfield  {author} {\bibinfo {author} {\bibfnamefont {Richard}\ \bibnamefont {Hill}},\ }\bibfield  {title} {\enquote {\bibinfo {title} {The elastic behaviour of a crystalline aggregate},}\ }\href@noop {} {\bibfield  {journal} {\bibinfo  {journal} {Proceedings of the Physical Society. Section A}\ }\textbf {\bibinfo {volume} {65}},\ \bibinfo {pages} {349} (\bibinfo {year} {1952})}\BibitemShut {NoStop}%
\bibitem [{\citenamefont {Jiang}\ and\ \citenamefont {Park}(2014)}]{Jiang_Nature_comm_2014}%
  \BibitemOpen
  \bibfield  {author} {\bibinfo {author} {\bibfnamefont {Jin-Wu}\ \bibnamefont {Jiang}}\ and\ \bibinfo {author} {\bibfnamefont {Harold~S}\ \bibnamefont {Park}},\ }\bibfield  {title} {\enquote {\bibinfo {title} {Negative poisson’s ratio in single-layer black phosphorus},}\ }\href@noop {} {\bibfield  {journal} {\bibinfo  {journal} {Nature communications}\ }\textbf {\bibinfo {volume} {5}},\ \bibinfo {pages} {4727} (\bibinfo {year} {2014})}\BibitemShut {NoStop}%
\bibitem [{\citenamefont {Anderson}(1963)}]{anderson_Jounal_of-phy_chem_of_solids_1963}%
  \BibitemOpen
  \bibfield  {author} {\bibinfo {author} {\bibfnamefont {Orson~L}\ \bibnamefont {Anderson}},\ }\bibfield  {title} {\enquote {\bibinfo {title} {A simplified method for calculating the debye temperature from elastic constants},}\ }\href@noop {} {\bibfield  {journal} {\bibinfo  {journal} {Journal of Physics and Chemistry of Solids}\ }\textbf {\bibinfo {volume} {24}},\ \bibinfo {pages} {909--917} (\bibinfo {year} {1963})}\BibitemShut {NoStop}%
\end{thebibliography}%

\clearpage
\widetext
\begin{center}
\textbf{\large Supplementary information for “Suppression of auxetic behavior in black phosphorus with sulfur substitution”}
\end{center}

\setcounter{equation}{0}
\setcounter{figure}{0}
\setcounter{table}{0}
\setcounter{page}{1}

\makeatletter

\renewcommand{\section}{\@startsection{section}{1}{0mm}
  {-\baselineskip}{0.0\baselineskip}{\bf\leftline}}
\renewcommand\thefigure{S\arabic{figure}}
\renewcommand\thetable{S\arabic{table}}
\renewcommand{\bibnumfmt}[1]{[S#1]}
\renewcommand*{\citenumfont}[1]{S#1}

\renewcommand{\thepage}{S\arabic{page}} 

\makeatletter
\def\blfootnote{\xdef\@thefnmark{}\@footnotetext}
\makeatother

\def\rev#1{\textcolor{RoyalBlue}{#1}}

\title{Supplementary information for ``Suppression of auxetic behavior in black phosphorus with sulfur substitution"}


\maketitle

\subsection{\label{sec:level1}Crystal structure}

Figure~\ref{fig: Supplemental_POSCARs} presents the DFT-optimized (PBEsol) crystal structures of pristine black phosphorus (b-P), two 25\% sulfur-substituted configurations (P$_6$S$_2$-II and P$_6$S$_2$-III, where two P atoms are replaced by two S atoms within the same phosphorene layer), and one with 50\% substitution (P$_4$S$_4$). For the 25\% substituted systems, although the Born–Huang mechanical stability criteria are satisfied, the structures undergo significant distortions.
In particular, the out-of-plane P–S bonds elongates significantly, disrupting the layered framework of b-P. This leads to pronounced changes in both lattice parameters and space group symmetries, as summarized in Table~\ref{tab:Supplemental lattice parameters}. On the other hand, the 50\% substituted system (P$_4$S$_4$) fails to satisfy the Born–Huang criteria, indicating mechanical instability.


\begin{figure*}[hbtp]
\centering
\includegraphics[width=0.85\textwidth]{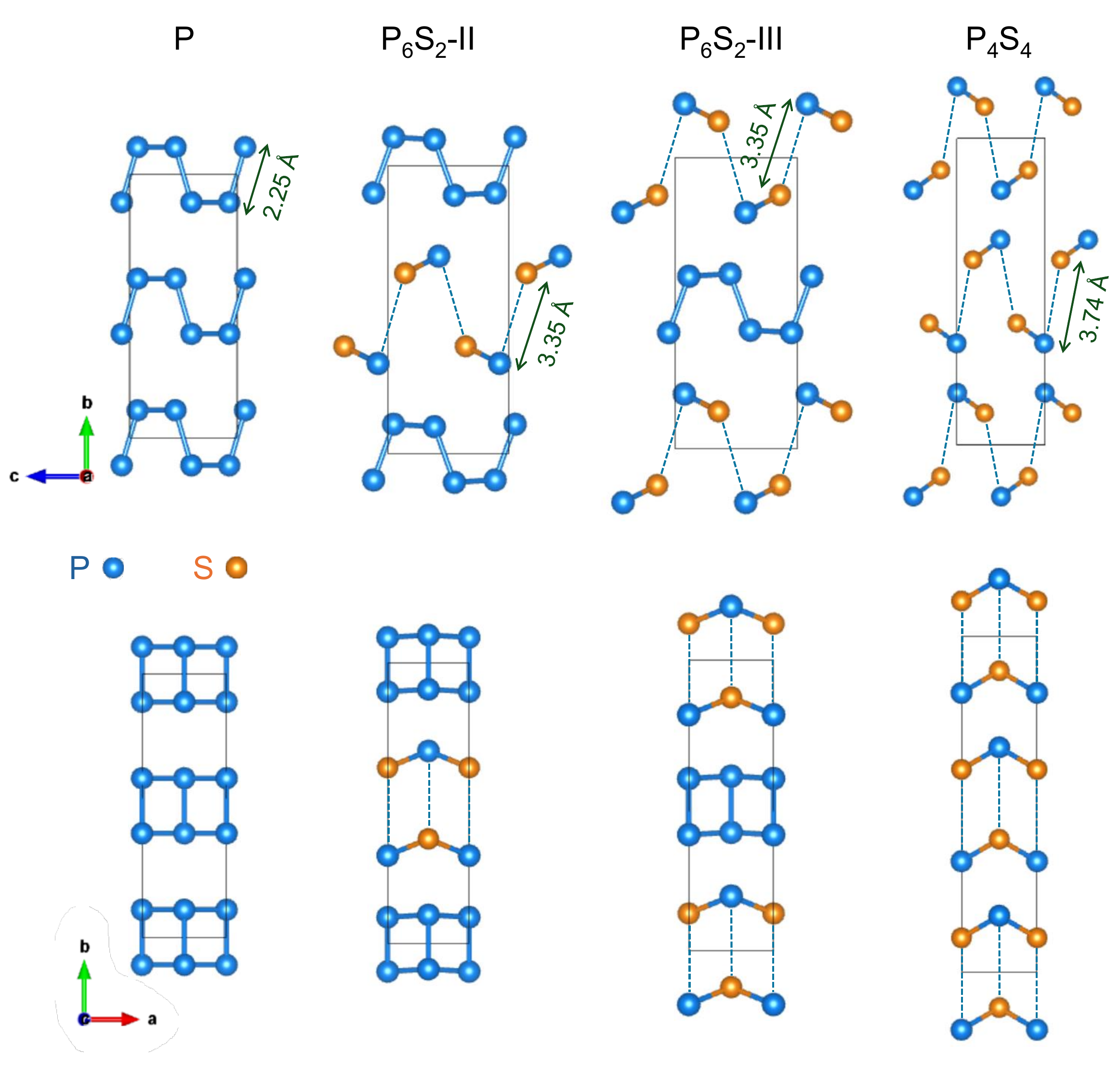}
\caption{Crystal structures of the P$_{1-x}$S$_{x}$ compositions omitted from the main text. Phosphorus atoms are shown in blue, and sulfur atoms are shown in orange. Dashed lines represent where a bond is expected but is not present in the DFT-optmized structures.}
\label{fig: Supplemental_POSCARs}
\end{figure*}

\begin{table}[hbtp]
\centering
\renewcommand*{\arraystretch}{1.2}
\caption{Lattice parameters and space groups of pristine black phosphorus (b-P) and P$_{1-x}$S$_x$ structures with 25–50\% sulfur substitution.}
\label{tab:Supplemental lattice parameters}
\begin{tabularx}{\textwidth}{l|*{4}{>{\centering\arraybackslash}X}}
\hline
Composition~~~& ~~ a (\r{A}) ~~&~~ b (\r{A})~~ & c (\r{A}) ~~ & ~~ Space Group\\
\hline\
b-P & 3.314 & 10.322 & 4.217 & Cmce (no.~64)\\
P$_6$S$_2$-II & 3.145 & 10.814 & 4.583 & Pmn$2_1$ (no.~31)\\
P$_6$S$_2$-III & 3.145 & 10.814 & 4.582 & Pmn2$_1$ (no.~31)\\
P$_4$S$_4$ & 3.017 & 13.456 & 3.849 & Pnma (no.~62)\\
\hline
\end{tabularx}
\end{table}

\subsection{\label{sec:level1}Elastic Properties}
Table~\ref{tab:supplemental elastic constants} summarizes the computed elastic constants (C$_{ij}$) for the two 25\% sulfur-substituted configurations (P$_6$S$_2$-II and P$_6$S$_2$-III). These systems exhibit anomalous behavior compared to the other S-substituted cases, which can be attributed to the significant structural distortions in their crystal frameworks. Consequently, this leads to substantial deviations in their elastic properties, as presented in Table~\ref{tab: Supplemental mechanical properties}. Figure~\ref{fig: Supplemental_ELFCARs} illustrates the distribution of electron localization function for pure black phosphorus, as well as for the two cases of sulfur substitution omitted from Figure 4 of the main text. 
We observe that the P$_{7}$S$_{1}$-II, as well as each individual layer of P$_{6}$S$_{2}$-I, are extremely similar in structure to the P$_{7}$S$_{1}$-I case. For this reason, they are omitted from the main text.


\begin{table*}[tbh]
\renewcommand*{\arraystretch}{1.25}
\caption{Elastic constants (C$_{ij}$, in GPa) of pristine black phosphorus (b-P) and P$_{1-x}$S$_x$ structures with 25–50\% sulfur substitution, calculated using PBEsol functional. The P$_4$S$_4$ case is omitted due to its mechanical instability.}

\label{tab:supplemental elastic constants}
\begin{tabularx}{\textwidth}{l|*{9}{>{\centering\arraybackslash}X}}
\hline
Composition~~~& ~~C$_{11}$ ~~&~~ C$_{22}$~~ & ~~C$_{33}$~~ & ~~C$_{44}$~~ & ~~C$_{55}$ ~~ & ~~C$_{66}$ ~~ & ~~ C$_{12}$ ~~ & ~~ C$_{13}$~~ & ~~ C$_{23}$~~\\
\hline\
b-P & 188.5 & 55.0 & 45.6 & 6.5 & 77.2 & 17.4 & 4.2 & 49.0 & -5.6\\
P$_6$S$_2$-II & 202.1 & 29.4 & 28.1 & 9.8 & 11.9 & 0.2 & -6.2 & 20.7 & 15.0\\
P$_6$S$_2$-III & 202.7 & 29.9 & 28.5 & 10.2 & 13.3 & 0.6 & -6.5 & 20.7 & 14.7\\

\hline
\end{tabularx}
\end{table*}

\begin{table*}[hbtp]
\renewcommand*{\arraystretch}{1.25}
\caption{Elastic properties of pristine black phosphorus (b-P) and P$_{1-x}$S$x$ structures with 25–50\% sulfur substitution, including bulk modulus K\,(GPa), shear modulus G\,(GPa), Young’s modulus E\,(GPa), average Poisson’s ratio $\nu$, longitudinal ($v_l$), transverse ($v_t$), and average ($v_m$) elastic wave velocities (m/s), as well as the Debye temperature ($\theta{_\text{Debye}}$) in K. The P$_4$S$_4$ structure is omitted due to its mechanical instability.}
\label{tab: Supplemental mechanical properties}
\begin{tabularx}{\textwidth}{l|*{8}{>{\centering\arraybackslash}X}}
\hline
Composition~~~& ~~ K (GPa) ~~&~~ G (GPa) ~~ & ~~E (GPa)~~ & ~~ $\nu$ ~~ & ~~$v_{l}$\,(km/s)~~ & ~~$v_{t}$\,(km/s)~~ & ~~ $v_{m}$\,(km/s)~~ & ~~ $\Theta_{\mathrm{Debye}}$\,(K)~~\\
\hline\
P & 32.3 & 26.2 & 61.8 & 0.187 & 4.855 & 3.030 & 3.338 & 379.0\\
P$_6$S$_2$-II & 28.1 & 10.4 & 26.5 & 0.370 & 3.970 & 1.976 & 2.218 & 245.4\\
P$_6$S$_2$-III & 28.1 & 11.3 & 28.9 & 0.353 & 4.026 & 2.061 & 2.309 & 255.5\\

\hline
\end{tabularx}
\end{table*}

\begin{figure*}[hbtp]
\centering
\includegraphics[width=0.7\textwidth]{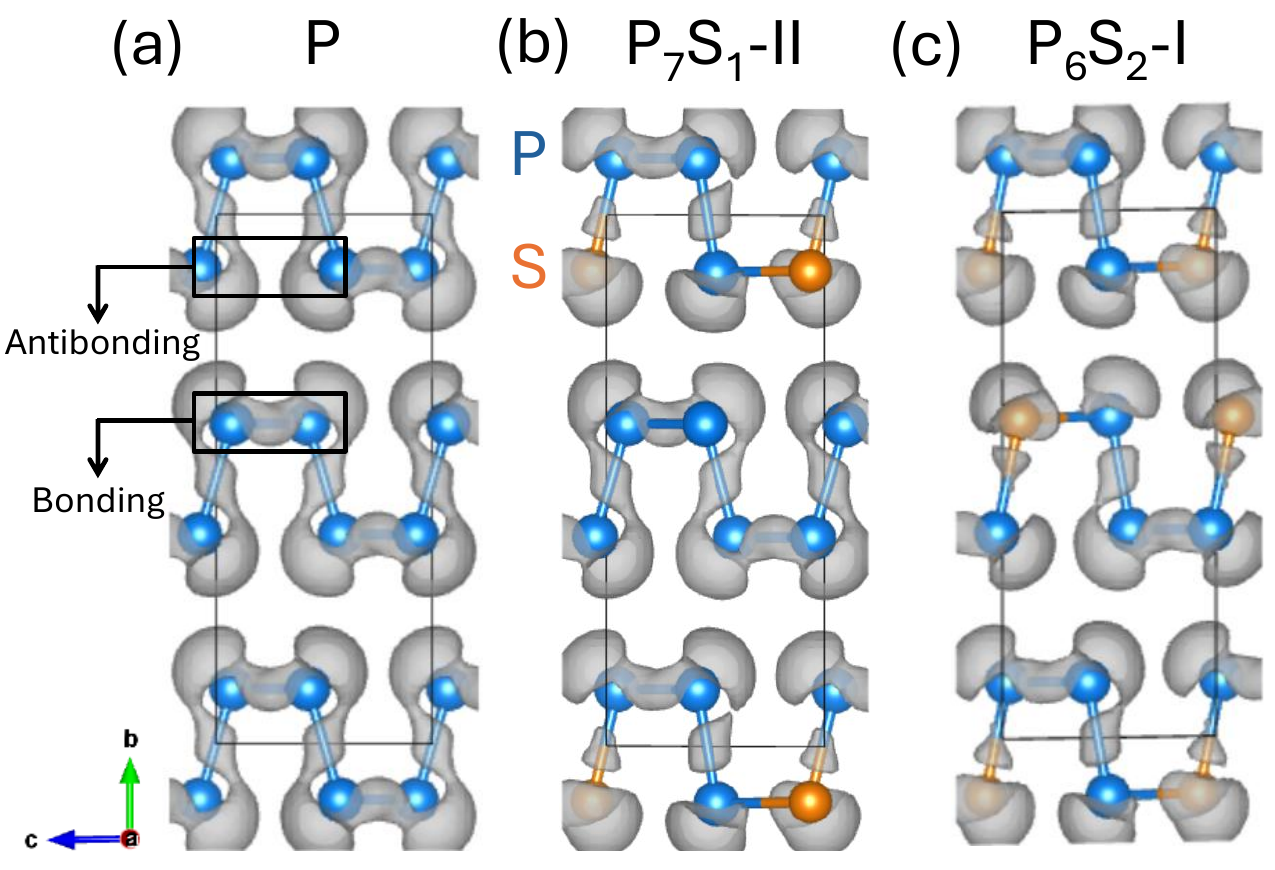}
\caption{Distribution of electron localization function
(gray) for the three cases of P$_{1-x}$S$_{x}$ that are omitted from Figure 4 of the main text (isosurface value = 0.8). Potential bonding and antibonding sites are marked in panel (a).}
\label{fig: Supplemental_ELFCARs}
\end{figure*}

\end{document}